%% file: dip1.tex
\documentstyle[aps,amstex,epsfig,subfigure]{revtex}
\title{Thermodynamics of superconducting lattice fermions}
\author{E. Otnes and A. Sudb{\o} }
\address{Department of Physics\\
	 Norwegian University of Science and Technology,
	 N-7034 Trondheim, Norway \\}

\begin{document}
\tightenlines
\maketitle

\begin{abstract}
We consider the Cooper-problem on a lattice model including onsite and 
near-neighbor interactions. Expanding the interaction in basis functions for 
the irreducible representation for the point group $C_{4v}$ yields a 
classification of the symmetry of the Cooper-pair wave function, which we 
calculate in real-space. A change of symmetry upon doping, from s-wave at low 
filling fractions, to $d_{x^2-y^2}$ at higher filling fractions, is found. 
Fermi-surface details are thus important for the symmetry of 
the superconducting wave function. Symmetry forbids mixing
of s-wave and d-wave symmetry in the Cooper-pair wavefunction on a square 
lattice, unless accidental degeneracies occur. This conclusion also holds
for the selfconsistent treatment of the many-body problem, at the 
critical temperature $T_c$. Below $T_c$, we find 
temperatures which are not critical points, where new superconducting channels 
open up in the order parameter due to bifurcations in the 
solutions of the nonlinear gap-equation. We calculate the 
free energy, entropy, coherence length, critical 
magnetic fields, and Ginzburg-Landau parameter $\kappa$. The  model is of the 
extreme type-II variety.
At the temperatures where subdominant channels 
condense, we find cusps in the internal energy and entropy, as well as
as BCS-like discontinuities in the specific heat. The specific heat
anomalies are however weaker than at the true superconducting critical 
point, and argued to be of a different nature. 
\end{abstract}

\section{Introduction}
The non-perturbative effect on the ground state wave-function  of an electron 
gas with arbitrarily weak effective attraction between the quasi-particles on 
the Fermi surface was first demonstrated by Cooper in his essentially exact 
solution of the corresponding two-body problem \cite{Cooper:1956}. Regardless of 
the strength of the interaction, two electrons interacting attractively 
on opposite sides of the Fermi-surface in an otherwise inert Fermi-gas form a 
bound state, the Cooper-pair. This simple calculation alone suffices to yield 
precisely the correct non-analytic dependence of the binding energy of the 
Cooper-pair on the dimensionless coupling constant $\lambda$ later found by 
solving the full selfconsistent problem \cite{Bardeen:1957}. Cooper solved 
the problem for the simple ``jellium" model of a metal, where the Fermi-sea 
was taken to be spherical, and the interaction between the two extra electrons 
of s-wave symmetry.  Of primary interest was the two-body spectrum, the 
Cooper-pair wave-function in k-space under such circumstances being a trivial 
constant. 

In this paper, we reconsider this simple problem on a square lattice.  The 
tight-binding band structure includes nearest and next-nearest neighbor hopping,
while the two-body term in the Hamiltonian includes  an onsite repulsive 
Hubbard-term, a nearest neighbor {\it effective} electrostatic interaction, 
which may be taken to be {\it attractive}, and also a next-nearest neighbor 
electrostatic interaction. The problem now includes two additional non-trivial 
features: i) The Fermi-surface is no longer spherical and one has to work 
directly in k-space rather than
transforming the problem to energy-space. ii) The interaction between the
quasiparticles no longer has simple s-wave symmetry, but rather may be 
expanded as a bilinear combination of basis functions for the irreducible
representation of the point group of the $2D$ square lattice, $C_{4v}$. This 
leads to the possibility of a number of interesting effects. In addition, we 
use a selfconsistent scheme to calculate the superconducting gap, thermodynamic 
quantities, and temperature dependence of critical magnetic fields of this 
phenomenological lattice fermion model.

This paper is organized as follows. In Section II, we define the model to
be considered. The method of calculating the Cooper-pair wavefunction is
presented in Section III, while specific numerical results pertaining to
this quantity are presented in Section IV. More detailed analytical results
on the binding energy of the Cooper-pair are given in Section V. In Section VI,
we present the results for  thermodynamic quantities and critical magnetic 
fields from a self-consistent scheme for a gap-function with several 
symmetry-channels, belonging to various s-wave and d-wave channels. 
In Section VII, we give a discussion of the specific heat anomalies
one may expect in such a model. 
We emphasize that throughout this paper, the superconducting order parameter 
is {\it not} a vector order parameter, but {\it assumed} to be a spin-singlet 
scalar complex order parameter such as is believed to describe conventional 
low-temperature superconductors and high-$T_c$ cuprates. 

\section{The model} 
The model we consider is an extended Hubbard-model on a square lattice
defined by the Hamiltonian
\begin{eqnarray}
H & = & - t \sum_{\langle i,j\rangle,\sigma} ~ c^+_{i,\sigma} ~ c_{j,\sigma} 
- t' \sum_{\langle\langle i,j\rangle\rangle,\sigma} ~ c^+_{i,\sigma} ~ c_{j,\sigma} 
- \mu  \sum_{i,\sigma} ~ c^+_{i,\sigma} ~ c_{i,\sigma}  \nonumber \\
 & + & \frac{1}{2} ~ 
\bigl[ 
\frac{U}{2} \sum_{i,\sigma} n_{i,\sigma} n_{i,-\sigma} 
+V 
\sum
\begin{Sb}
\langle i,j\rangle, \\
\sigma,\sigma'
\end{Sb}
 n_{i,\sigma} n_{j,\sigma'} 
+W
\sum
\begin{Sb}
\langle\langle i,j\rangle\rangle \\
\sigma,\sigma'
\end{Sb}
n_{i,\sigma}n_{j,\sigma'}
\bigr].
\label{Hamiltonian}
\end{eqnarray}
Here $\langle i,j\rangle$ and $\langle\langle i,j\rangle\rangle$ denote nearest 
neighbor and next-nearest neighbor couplings, respectively. $t$, and $t'$ are 
corresponding hopping matrix elements, and $\mu$ is the chemical potential. $U$ 
is an onsite repulsion term, while $V$ and $W$ are effective electrostatic 
Coulomb 
matrix-elements between nearest and next-nearest neighbors, respectively. When 
viewed as an effective interaction term within a one-band model, as reduced from 
a multiband-band model, {\it these terms may be attractive}
\cite{Varma:1987,Emery:1987,Stechel:1995,Varma:1997}. In this 
paper we simply take them
as effective attractions without further justification.

After introducing a plane-wave basis and performing a standard BCS truncation
of the interaction piece of the Hamiltonian \cite{Bardeen:1957,AGD:1980}, 
it takes the usual form
\begin{equation}
H = \sum_{ \vec k,\sigma} ~ \varepsilon_{\vec k} 
~ c^+_{\vec k,\sigma} ~ c_{\vec k,\sigma} 
+ \sum_{\vec k, \vec k'} ~ V_{\vec k,\vec k'} 
~ c^+_{\vec k,\uparrow}   ~ c^+_{-\vec k,\downarrow} 
~ c_{-\vec k',\downarrow} ~ c_{\vec k',\uparrow}, 
\end{equation}
where we have defined, after an appropriate redefinition of the zero-point 
of energy and a rescaling of $t'$ and $\mu$
\begin{equation}
\varepsilon_{\vec k} = - 2 t \bigl[ \cos(k_x) + \cos(k_y) 
                       - 2 t' \cos(k_x) \cos(k_y) -(2-2t'-\mu) \bigr].
\label{free-spectrum}
\end{equation}
The choice of such a quasiparticle dispersion is obviously motivated by its 
relevance as a simple means of modeling the quasi-particle band crossing the 
Fermi-surface of the high-$T_c$ cuprates \cite{Freeman:1991}. Note that 
in such a context, the inclusion of the $t'$-term is crucial; a bipartite 
lattice with nearest-neighbor hopping only, is inconsistent with the observed 
Fermi-surfaces in the high-$T_c$ cuprates, where $|t'| \approx |t|/2$ 
\cite{Shen:1993}. (This also has motivated the choice $t'=0.45 t$ in our 
numerical calculations).  The importance
of including this term in correctly interpreting experiments, has recently
been strongly emphasized \cite{Ganpathy:1997}. 

In the above truncation of the interaction term, 
the interaction is assumed to be operative between 
fermion spin-singlets on opposite sides of the Fermi-surface, the inert and 
rigid Fermi-sea merely giving rise to Pauli-blocking factors. With the 
interactions  given in Eq. \ref{Hamiltonian}, it is readily shown that
$V_{\vec k,\vec k'}$ is given by
\begin{equation}
V_{\vec k,\vec k'} = \sum_{\eta=1}^5 ~ 
\lambda_{\eta} ~ B_{\eta} (\vec k) ~ B_{\eta} (\vec k'),
\label{potential}
\end{equation}
where $\lambda_1=U/2$, and $\lambda_2=\lambda_4=V$, $\lambda_3=\lambda_5=W$. 
We  have also found it convenient to introduce the simple, but sufficient, 
subset of basis functions $\{ B_{\eta}(\vec k) \}$ for irreducible 
representations of the symmetry group $C_{4v}$ of the square lattice, 
$B_1(\vec k)  =  \frac{1}{\sqrt{N}}$, 
$B_2(\vec k)  =  \frac{1}{\sqrt{N}}[\cos(k_x) + \cos(k_y)]$,
$B_3(\vec k)  =  \frac{2}{\sqrt{N}}[\cos(k_x) ~ \cos(k_y)]$,
$B_4(\vec k)  =  \frac{1}{\sqrt{N}}[\cos(k_x) - \cos(k_y)]$,
$B_5(\vec k)  =  \frac{2}{\sqrt{N}}[\sin(k_x) ~ \sin(k_y)]$, where
$N$ is the number of lattice sites. Inclusion of longer ranged interactions 
will in general require an augmentation of this subset, but any finite
ranged interaction will yield a separable potential.

\section{The Cooper pair wave function}

We define a two-particle state for the non-interacting case, i.e. $U=V=W=0$, 
by $|k,\sigma;-k,-\sigma{\rangle_0}$ obeying the Schr{\"o}dinger equation 
\begin{equation}
H_0|k,\sigma;-k,-\sigma\rangle_{0} 
= 2\varepsilon_{\vec{k}}|k,\sigma;-k,-\sigma ~\rangle_{0},
\end{equation}
where $H_0$ denotes the Hamiltonian of the free particles, and 
$\varepsilon_{\vec k}$ is given in Eq. \ref{free-spectrum}. Note that in 
this notation, the hopping between next nearest neighbors is given by the 
matrix element, $4t*t'$, and we limit ourselves to situations where $2 |t'|<1$, 
such that the bottom of the band $\varepsilon_{\vec k}$ is located at the 
Brillouin-zone center. 

The problem we will consider is a simplification of the one posed by 
$H=H_0+H_{\rm{int}}$. We imagine that we have a rigid Fermi-sea of 
non-interacting electrons with a spectrum given by Eq. \ref{free-spectrum}.  
To this inert Fermi-sea we add two electrons on opposite sides  of the 
Fermi-sea which interact with the matrix element $V_{k,k'}$. This interaction 
term scatters a pair of electrons in the state $|k,\sigma;-k,-\sigma\rangle_0$ 
to the state $|k',\sigma;-k',-\sigma\rangle_0$. The exact two-particle state 
for the two extra electrons, for which $\vec k$ is no longer a good quantum 
number, is defined by $|1,2\rangle$. This state is expanded in two-particle 
plane-wave states as follows
\begin{equation}
|1,2 {\rangle}=
\sum_{k>k_F,\sigma} ~ 
a_{\vec{k},\sigma} ~ |k,\sigma;-k,-\sigma \rangle_{0},
\end{equation}
such that the problem is reduced to one of determining the Fourier-coefficients
$a_{\vec k,\sigma}$. The exact two-particle state for this problem obeys the 
Schr{\"o}dinger equation:
\begin{eqnarray}
(H_0+H_{\rm{int}})|1,2 {\rangle}& = & E|1,2 {\rangle}.  \nonumber
\end{eqnarray}
Upon inserting the expansion in plane-wave states, and projecting onto plane-wave
states, we obtain  upon using Eq. \ref{potential}, an integral equation for 
the expansion coefficients 
\begin{equation}
\sum_{\eta}B_{\eta}(\vec{k'})\underbrace{\sum_{k>k_F}
~ \lambda_{\eta} ~ a_{\vec{k}}  
B_{\eta}(\vec{k})}_{A_{\eta}} = (E - 2\varepsilon_{\vec{k'}})
~ a_{\vec{k'}}, \nonumber   
\end{equation}
which immediately yields an expression for the expansion coefficients
\begin{equation} 
a_{\vec{k'}}  =  \frac{\sum_{\eta} A_{\eta}B_{\eta}(\vec{k'})}
{E - 2\varepsilon_{\vec{k'}}};\qquad  \varepsilon_k>\varepsilon_{k_F}.
\end{equation}
Here $B_{\eta}(\vec{k})$ is one of the five basis functions required to
expand the interaction.  Finding the wave function, $a_{\vec{k}}$ thus amounts 
to finding the eigenvalue $E$ and the amplitudes $A_{\eta}$. Inserting the 
expression obtained above for $a_{\vec k}$ and $V_{\vec k \vec k'}$
into the Schr{\"o}dinger equation, we obtain coupled 
algebraic equations for the amplitudes, $A_{\eta}$ as follows 
\begin{align}
 \sum_{\eta'}\lambda_{\eta'} B_{\eta'}(\vec{k'})\sum_{\eta} A_{\eta}
\underbrace{\sum_{k>k_F} \frac{B_{\eta}(\vec{k})B_{\eta'}(\vec{k})}{E -
2\varepsilon_{\vec{k}}}}
_{D_{\eta\eta'}}
&=
\sum_{\eta'}A_{\eta'} B_{\eta'}(\vec{k'}).              
\end{align}
The functions $B_{\eta}(\vec{k})$ are linearly independent, and 
a comparison of coefficients therefore yields 
the following set of  equations for the amplitudes $A_{\eta}$
\begin{eqnarray}
\sum_{\eta} A_{\eta} ~ \lambda_{\eta'} ~ D_{\eta \eta'} 
= A_{\eta'}. 
\label{aeqs} 
\end{eqnarray}
The result can be written as a vector equation,
$\mathbf{(T - I)}\vec{A}=0$,
where $T_{\eta'\eta}=\lambda_{\eta'}D_{\eta\eta'}$ and $\mathbf{I}$ denotes
the identity matrix. A nontrivial solution exists if and only if the system 
determinant vanishes $|\mathbf{T - I}| = 0$, which 
determines the eigenvalue $E$,  and thus in turn the eigenvectors.

Note that this analysis shows that at this level, in general one cannot 
obtain a Cooper-pair wave function with a mixed $(s,d)$-symmetry.
This is a consequence of the fact that the matrix $D_{\eta \eta'}$
is block-diagonal in the s-wave and d-wave sectors due to the fact that
the ``pair-susceptibility" $\chi_k = 1/(E-2 \varepsilon_{\vec k})$ transforms
as a function with s-wave symmetry expandable in the functions $B_1$, $B_2$ and 
$B_3$. In Section VI, we show that this conclusion holds for the mean-field 
gap-equation that results from a self-consistent solution to the full problem 
at the critical point $T_c$. The wave-function may thus have one of the 
following forms
\begin{eqnarray}
a_{\vec{k}}
&= & \sum_{\eta=1}^3
 ~ A_{\eta} ~ B_{\eta}(\vec{k}) ~ \chi_{\vec{k}}~;~~~~
\varepsilon_k>\varepsilon_{k_F},  \nonumber \\
a_{\vec{k}}	
&= & \sum_{\eta=4}^5 ~ A_{\eta} ~ B_{\eta}(\vec{k}) ~ \chi_{\vec{k}}~;
~~~\varepsilon_k>\varepsilon_{k_F}. 
\end{eqnarray}
All the amplitudes $a_{\vec k}$ are zero for $\varepsilon_{\vec k} < 
\varepsilon_{k_F}$. Since the $\mathbf{D}$-matrix is block diagonal, we may 
find two different eigenvalues, one for each irreducible representation. The 
correct eigenvector corresponds to the lowest eigenvalue $E$. Furthermore, an 
immediate consequence of the fact that 
$A_\eta=\sum_{k>k_F} ~ \lambda_{\eta} ~ a_k  ~ B_{\eta}(k)$
is that $A_{\eta}\neq 0$ if and only if $\lambda_{\eta}\neq0$.
Specifying the $\lambda_{\eta}$'s thus immediately determines 
which of the basis functions $B_{\eta}(\vec{k})$ that
may contribute to the Cooper pair wave function.

The Cooper pair wave function in real-space is calculated by applying the 
inverse lattice Fourier-transform to $a_{\vec k,\sigma}$, defined by
\begin{equation}\label{invFT}
\psi(\vec{r})=\frac{1}{N}\sum_{k_1=0}^{n_k-1}\sum_{k_2=0}^{n_k-1}
a_{\vec{k}}e^{i\frac{2\pi}{n_k}
({\langle}k_1,k_2{\rangle}{\langle}i,j{\rangle})},
\end{equation}
where $\vec{r}={\langle}i,j{\rangle}$ is the relative distance between the 
electrons comprising the Cooper-pair, $\vec{k}={\langle}k_1,k_2{\rangle}$.

\section{Numerical Results} 

In Fig. \ref{fig:U0muchange} (a)-(f) we have plotted $|\psi(i,j)|^2$
of the Cooper pair wave function given by Eq. (\ref{invFT}) in the
case of no onsite interaction, $U=0$, and nearest neighbor attraction, 
$V=-0.75*t$. Throughout this discussion, we set $W=0$. The hopping matrix 
element $t=0.10\,eV$. At low filling fractions, the wave function displays 
s-wave symmetry which can be recognized by the peak present at the  point 
$(0,0)$. Electrons situated at the same site forming a pair cannot be in a 
d-wave state. An example of a d-wave state is shown in  
Fig. \ref{fig:U0muchange}(c). This state is in fact found when increasing the 
filling fraction to $n=0.17$, for the same value of the Coulomb-parameters. 
The pairing state has changed its transformation properties due to doping. 
Further increasing the filling does not change the symmetry, but Figs. 
\ref{fig:U0muchange}(c)-(f) show that the pairing state broadens as doping is 
increased. The region where  $|\psi(i,j)|^2 \ne 0$ can roughly be interpreted 
as the size of the Cooper pair, i.e. the coherence length, which is thus seen 
to increase with increasing filling fraction. This is verified directly by 
calculating the quantity 
$\sum_{i,j} ~ |i-j|^2 ~ |\psi(i,j)|^2/\sum_{i,j} |\psi(i,j)|^2$. 

To further investigate at which doping level the symmetry change occurs, we 
have plotted the eigenvalue $E$ of the two-particle Schr{\"o}dinger equation, 
in Fig. \ref{fig:enrgU0} as a function of doping. The eigenvalue $E$ is 
determined by numerically solving $|{\bf T}-{\bf I}|=0$. The Fig. shows that 
for small filling fractions, i.e. $n<0.1$, the eigenvalue for s-wave pairing 
is lower than the corresponding value for d-wave pairing, and hence s-wave 
pairing is therefore energetically favorable. At $n \approx 0.10$, we 
see that the two energy curves intersect, and for $n>0.10$,  d-wave pairing is
thus expected to be energetically favorable.

We have done similar eigenvalue calculations as above in the case of an onsite 
repulsion, $U=1.0*t$. The results are shown in Fig. \ref{fig:enrgU1}. Since 
an onsite interaction does not affect d-wave pairing, the curve that shows the 
d-wave eigenvalues is the same as shown in the scenario of no onsite 
interaction. The s-wave curve has however changed, and appears to be shifted 
to higher energies, which means that the change in the symmetry of the 
wave-function occurs at a lower filling fraction, in this case $n\approx0.08$. 
The results presented in Fig. \ref{fig:enrgU1} indicate that increasing
onsite repulsion favors d-wave pairing, as one would expect. The 
extended s-wave component $B_2$ of the Cooper-pair wave function 
will inevitably have a finite on-site component, although it avoids the hard 
core to a considerable extent. 

Figs. \ref{fig:U4muchange}(a)-(f) show the results for $|\psi(i,j)|^2$  as a 
function of doping for $U=4.0*t$.  This onsite repulsion is large enough to 
suppress pairing in any of the s-wave channels for all fillings that we have 
considered, in the range $n \in[0.06,0.85]$.  As filling is increased, one 
starts 
seeing a considerable broadening of the wavefunction with increasing filling
fraction.

Figs. \ref{fig:mu002changeU} (a)-(f) show the 
Cooper-pair wave function with $n=0.06$ at six different values for the 
onsite repulsion $U$. In Fig. \ref{fig:mu002changeU}(a), the  onsite repulsion 
$U=0$, and the s-wave solution is the energetically favorable (see
Fig. \ref{fig:enrgU0}). In Fig. \ref{fig:mu002changeU}(b) a {\it weak} onsite 
repulsion $U=0.5*t$ is applied. Note that for this case, we obtain the somewhat 
counter-intuitive result that, since $|\psi(i,j)|^2;~~|i-j|=0$ evidently 
has increased 
upon increasing the onsite repulsion,  the weak increased repulsion effectively 
promotes attraction between electrons. Similar effects have been seen in exact 
many-body calculations on strongly correlated $1D$ lattice fermion models 
\cite{Sandvik:1996}.  We will comment on this result later. 
Figs. \ref{fig:mu002changeU}(c)-(f) display the wave function when the onsite 
repulsion is increased in the range $U/t =(0.7,1.0,1.30,1.45)$ and it is 
evident that between $U=1.30*t$ and $U=1.45*t$, the wave 
function has changed its transformation properties to d-wave in order to 
completely eliminate the effect of the hard core. Increasing the onsite 
repulsion further will not affect the d-wave function, as already mentioned.

To show in more detail for which values of $U$  the wave function changes its 
transformation properties, we have calculated the energy eigenvalues as a 
function of onsite repulsion  for a filling fractions $n=0.06$. 
The results are shown in Fig. \ref{fig:enrgmu002}.
The trends 
in the results clearly show that in lattice models with four-fold symmetry, 
s-wave superconducting pairing is a low-filling effect, while d-wave pairing 
practically always wins in situations close to half-filling.

\section{Analytical Results}
It is instructive also to perform a simplified analytical treatment of the
Cooper-problem. A major simplification results by rewriting 
$D_{\eta \eta'}(E)$ as follows
\begin{eqnarray} \label{Eff_dos}
D_{\eta \eta'}(E) & = & 
\int_{-\infty}^{\infty} ~ d \varepsilon \sum_{\vec k}
\delta(\varepsilon - \varepsilon_{\vec k} )  ~ \theta(\varepsilon_{\vec k})
~ \frac{B_{\eta}({\vec k}) ~ B_{\eta'}({\vec k})}
     {E-2 \varepsilon} 
 =  \int ~ d \varepsilon \frac{N_{\eta \eta'} (\varepsilon)}
                                {E - 2 \varepsilon}  ~
\theta(\varepsilon) \nonumber \\
& \approx & {N_{\eta \eta'}} ~ \int_{\mu}^{\omega_c+\mu} d \varepsilon
~ \frac{1}{E-2 \varepsilon} = 
{N_{\eta \eta'}} F(\frac{\Delta}{\omega_c}),
\label{approx} 
\end{eqnarray}
where we have introduced the binding-energy 
$\Delta \equiv 2 \mu - E$, $\omega_c$ is an upper band-cutoff and the function 
$F$ is defined as $F(x) \equiv \frac{1}{2} ~ \ln(1 + \frac{2}{x})$. In addition, 
we have introduced the projected  densities of state 
$N_{\eta \eta'}(\varepsilon)  =  \sum_{{\vec k}} ~ 
~~ B_{\eta}({\vec k}) ~ B_{\eta'}({\vec k}) ~ 
\delta(\varepsilon - \varepsilon_{\vec k})$. The constants $N_{\eta \eta'}$ in Eq. \ref{approx} are chosen appropriately to 
get the correct value for the $\vec k$-space sum. With our choice of coupling 
constants, the matrix $D_{\eta \eta'}$  block-diagonalizes  into a $3 \times 3$ 
matrix in the s-wave sector, and a $2 \times 2$ matrix in the d-wave sector. 
For simplicity, in the following we will set $W=0$. It will also be convenient 
to introduce the following definitions
\begin{eqnarray}
\nu &\equiv & \frac{UV}{2} ~
({N_{11}} ~~ {N_{22}} - {N_{12}}^2);~~~~ 
\gamma \equiv  \frac{1}{2} ~ (\frac{U}{2} {N_{11}} + 
V {N_{22}}). 
\end{eqnarray}
With these definitions, we get the following secular equation for $F$ (and hence 
$\Delta$) in the s-wave sector $\nu ~ F^2 + 2 \gamma ~ F + 1 = 0$. 
This is a necessary, but not sufficient, condition for finding a 
non-zero Cooper-pair wave function in the s-wave sector. Note that we must 
in principle distinguish between four cases: 
i) $\nu < 0, ~~~~\gamma>0$,  
ii) $\nu < 0, ~~~~\gamma<0$, 
iii) $\nu > 0, ~~~~\gamma>0$,  
iv) $\nu > 0, ~~~~\gamma<0$. 
Here, we will discuss the case $\nu < 0$, which should be realized for
appropriate band-fillings.  
Furthermore, it is clear that in the large-$U$ limit, $\gamma > 0$, while 
$\gamma < 0$ when $U=0$. Hence, we have 
\begin{eqnarray}
F & = & \frac{|\gamma| + \sqrt{|\gamma|^2+|\nu|^2}}{ |\nu|} = \frac{1}{2 \lambda_s}; ~~~ 
U > \frac{|V| ~~{N_{22}}}{2} \nonumber \\
F & = & \frac{-|\gamma| + \sqrt{|\gamma|^2+|\nu|^2}}{ |\nu|} = 
\frac{1}{2 \lambda_s}; ~~~ 
U < \frac{|V| ~~{N_{22}}}{2},
\label{thefs}
\end{eqnarray} 
which defines the dimensionless coupling constant $\lambda_s$. It follows that 
when $U=0$, we get $F= 2/|V| ~~ N_{22}$ while for $U=\infty$, we find
$F= 2/(|V| ~~ (N_{22}-N_{21}^2/N_{11})$.  Note that, within the s-wave 
sector, even an infinite hard-core repulsion does not suffice to destroy 
superconductivity. This is mainly due to the fact that 
the extended s-wave piece of the wave-function avoids the hard core. What will 
ultimately determine the symmetry of the superconducting wave-function is 
therefore the competition with the d-wave sector. In the d-wave sector, we 
obtain the secular equation $1-2 \lambda_d ~ F = 0$. \\
In either case, to have a bound state form, it is required that  
$F$ is real  and positive, equivalently the effective dimensionless coupling 
constants $\lambda_s$ and $\lambda_d$  must be positive. 
The only acceptable solution to the secular equation in the s-wave sector is 
$F   =  \frac{1}{2 \lambda_s};~~~~  
\Delta_s  =  \frac{2 \omega_c}{\exp(1/\lambda_s)-1}$,
while for  the d-wave sector we obtain 
$F  =  \frac{1}{2 \lambda_d};~~~~ 
\Delta_d  =  \frac{2 \omega_c}{\exp(1/\lambda_d)-1}$.
The binding energy, in the s-wave and d-wave sectors, are determined
by the effective coupling constants $\lambda_s$ and $\lambda_d$, respectively.
The system chooses the symmetry of the superconducting wave-function which 
gives the largest binding energy, equivalently the largest 
$\lambda$. \\
To gain some insight into what filling fractions one might expect one or the 
other coupling constant to dominate, it is instructive to exhibit 
$N_{22}(\varepsilon)$ and $N_{44}(\varepsilon)$ as a function 
of the energy. They are shown in Fig. \ref{fig:ProDos}, along with the 
single-particle density of states $N(\varepsilon)=\sum_{\vec k} 
\delta(\varepsilon-\varepsilon_{\vec k})$,  which we have been able to evaluate
analytically. This is useful since it provides intuition about
the singularities that also appear in the projected densities of states,
and which we have to consider when the stability of the various 
superconducting channels are investigated. Using Eq. \ref{free-spectrum}, we 
find straightforwardly that the single-particle density 
of states is given by   \cite{Xing:1991}
\begin{eqnarray}
N(\varepsilon)  & = &
\frac{1}{2 ~ \pi^2 ~ |t|} ~ \frac{1}{\sqrt{1-\rho ~  a}}
~ {\bf K} \Bigl( \sqrt{\frac{1-[(a + \rho)/2]^2}{1-\rho ~ a}} 
\Bigr);~~~  |(a + \rho)/2| < 1 \nonumber \\
& \approx & 
\frac{1}{2 ~ \pi^2 ~ |t|} ~ \frac{1}{\sqrt{1-\rho ~  a}}
~ \ln \Bigl( \frac{8 ~ \sqrt{1-\rho ~ a}} {|a-\rho|} \Bigr);~~~  
|(a + \rho)/2| < 1 \nonumber \\
       & = & 0; ~~~ |(a+\rho)/2|>1,
\label{andos}
\end{eqnarray}
where we have defined $\rho=-2t'$, $a=\varepsilon/2t$, and ${\bf K}(k)$ is a 
complete elliptic integral of the first kind, with modulus 
$k$ \cite{Gradshteyn}. The result is valid for $|\rho| < 1$, the 
requirement for having a band-minimum at the center of Brillouin-zone. 
(It is quite remarkable that the approximate logarithmic expression in 
Eq. \ref{andos} reproduces the exact expression excellently over the entire 
band. The reason is that the main effect of the elliptic integral ${\bf K}$ 
is to produce a narrow logarithmic singularity at $a=\rho$, while the 
remaining variation in $N(\varepsilon)$ is due to the prefactor.) The density 
of states has a logarithmic singularity $N(\varepsilon) = 
(1/2 \pi^2 |t| \sqrt{1-\rho^2}) ~ \bigl[ \ln (8/|a-\rho|) + 
\ln(\sqrt{1-\rho^2}) \bigr]$ at $\varepsilon = 2 ~ t ~\rho$, a finite cusp at 
the lower band-edge, $N(\varepsilon=2 t(-2-\rho))=(1/4 \pi |t|)/(1+\rho)$,  
while the density 
of states close to the upper band-edge has a smooth behavior with a value at 
the upper band-edge given by $N(\varepsilon=2 t(2-\rho))=
(1/4 \pi |t|)/(1-\rho)$. Were we to
reverse the sign of $t'$, we would get a cusp at the upper band-edge, while 
the density of states at the lower edge would be smooth. The singularities
that appear in the projected densities of states must come from the
singularities in $N(\varepsilon)$. 
We have not been able to reduce the integrals for the projected 
densities of states
to useful expressions in terms of elliptic integrals, and have simply
calculated them numerically using the tetrahedron algorithm
\cite{Lehmann:1972}. \\
The peaks in $N_{22}(\varepsilon)$  originate from energies close to the bottom 
of the band, essentially identified with the cusps in $N(\varepsilon)$, whereas
the peaks in the $N_{44}(\varepsilon)$ come from energies identified with
the logarithmic singularities in $N(\varepsilon)$. As soon as the filling 
fraction increases from the bottom of the band, the major singularities 
in $N_{22}(\varepsilon)$ are eliminated from the integrated
projected density of states $N_{22}$, and the contributions to $N_{44}$ coming
from the logarithmic singularities in $N_{44}(\varepsilon)$ at 
$\varepsilon = 2 t \rho$ will 
dominate. Hence, $\lambda_d$ and thus d-wave pairing will always dominate 
close to half-filling, while s-wave pairing will win out close to the bottom 
of the band. Note for instance in the case where $t'=0$,
that $N_{22}(\varepsilon)=0$ in the middle of the band, while 
$N_{44}(\varepsilon)$ has a weak
logarithmic singularity. When $t'\neq 0$ the situation changes somewhat,
and both projected densities of state are finite at half filling. 
Nonetheless, as long as $t'<0.5$, such that the bottom of the band
is located at the zone-center, $N_{44} > N_{22}$ for filling fractions 
such that the Fermi-surface is close to the middle fo the band.  For 
non-pathological situations, d-wave pairing will in general dominate close 
to half-filling. The conclusion holds for any pairing kernel with a 
predominant $B_{4}$-symmetry.\\
A somewhat counter-intuitive result obtains if we 
introduce a weak repulsive coupling $U$ in the problem. In our notation, this 
means that in Eq. \ref{thefs}, we may expand $F$ in Eq. \ref{thefs} in powers 
of $|\nu|/|\gamma|$, in order to find the corrections to $\lambda_s$ to 
leading order in $U/V$. The leading order correction in U is given by
\begin{eqnarray}
\lambda_s - |\gamma| = \frac{|\nu|}{4~ |\gamma|} \sim |U|.
\end{eqnarray}
Thus, we see that an {\it enhancement} of $\lambda_s$, and hence an enhancement 
of superconductivity, comes about initially as $U$ is switched on, 
{\it irrespective of the sign of $U$}. On the other hand, if we consider 
$\lambda_s$ at $U=\infty$, it is clearly less than $\lambda_s$ at $U=0$,
\begin{eqnarray}
\lambda_s(U=0)-\lambda_s(U=\infty) = \frac{|V|}{2} ~~ 
\frac{{N_{12}}^2}{{N_{11}}} > 0.
\end{eqnarray}
It follows that there must exist an intermediate value of $U$ which maximizes 
the binding energy in the s-wave sector. Since $\lambda_d$ is independent
of $U$ altogether, we obtain the counter-intuitive result that there exists
a regime of filling-fractions, where one may obtain a ``switch" of the
symmetry of the superconducting wave function from d-wave to extended s-wave 
upon increasing a weak onsite repulsion. 
This has much in common  with the result we observed in the numerics, that 
an increase in $U$ for $n=0.06$ produced 
a more tightly bound Cooper-pair. 
While the analytical results quoted above are approximate, the
numerical results are quite remarkable in that they, within the
Cooper-scheme, entail no further approximations. We also note 
that similar results have been found for the 
zero-temperature superconducting gap in self-consistent mean-field calculations 
\cite{O'Donovan:1995}. We now comment on why we may obtain such a 
counter-intuitive result in our numerical calculations. 

A finite interaction $V < 0$  implies that when $U=0$, there certainly will
exist a solution to the Schr{\"o}dinger equation such that 
$A_{2} \neq 0$, i.e. we are guaranteed pairing in the extended s-wave channel 
{\it for low enough band-fillings}, as argued above based on our intuition on 
the projected densities of states. Increasing $U$ slightly from zero, implies 
that one in principle is coupling in another component of the wave-function, 
$A_1$, since $A_1 \sim \lambda_1 \sim U$. However, $A_1$ and $A_2$ may enter 
in the wavefunction with relative phases of $\pi$. The coupling between $A_1$ and 
$A_2$ in the gap-equation may take advantage of this and essentially produce an
``attraction from a repulsion" by appropriately twisting the relative phases 
of the amplitudes entering in the order parameter. Note that such a mechanism
cannot work with an order parameter containing only one symmetry-channel 
component, since the overall phase of the wavefunction is irrelevant for the
spectrum. In the 
present situation with two s-wave components, however, the extended s-wave 
amplitude is able, for small values of $U$, to ``drag" the isotropic s-wave
component of the order parameter along, as a consequence of the 
``conversion of repulsion to attraction"-effect. Continuing to increase the
onsite repulsion to larger values $U/t >>1$, energetics dictates that  
the on-site component of the order parameter eventually must vanish, thereby 
reducing the indirect attractive effect of the onsite-repulsion. 
As a consequence, the pair-wave function will again spread out in real
space, as observed in our numerics. 

Note that the above results are not sensitive to the basic pairing 
mechanism operative in the underlying microscopic theory. All that enters 
at this level, are competing channels in the pairing kernel with isotropic 
and extended s-wave symmetries, as well as d-wave symmetries.
Hence, for instance anti-ferromagnetic spin-fluctuations
\cite{Scalapino:1992}, who some believe to be relevant
in producing superconductivity in the high-$T_c$ cuprate
$CuO_2$-planes, would fit into the scenario 
Eq. \ref{potential} for the pairing kernel. Nonetheless, 
the above ``attraction from repulsion" effect within the 
isotropic and extended s-wave sector is not particularly
relevant to  the physics of the high-$T_c$ oxides, which show
superconductivity only in the vicinity of half-filled bands, where
we find that $d_{x^2-y^2}$-wave pairing almost invariably wins out.

\section{Selfconsistent Analysis}
In this Section, we consider the full nonlinear mean-field gap-equation. 
Starting from the full Hamiltonian, Eq. \ref{Hamiltonian}, one performs a 
standard BCS-truncation of the interaction term, and a further anomalous
mean-field decomposition to obtain the gap-equation  
\begin{eqnarray}
\Delta_{\vec k} & = & -\sum_{\vec k'} ~ V_{\vec k,\vec k'} ~ 
\Delta_{\vec k'} ~  \chi_{\vec k'} \nonumber \\
V_{\vec k,\vec k'} & = & \sum_{\eta} ~ 
\lambda_{\eta} ~ B_{\eta}(\vec k) ~ B_{\eta}(\vec k')  \nonumber \\
\chi_{\vec k} & = & \frac{1}{2 E_{\vec k}} ~ \tanh \Bigl(
\frac{\beta ~ E_{\vec k}}{2} \Bigr) \nonumber \\
E_{\vec k} & = & \sqrt{\varepsilon_{\vec k}^2 + |\Delta_{\vec k}|^2}.
\end{eqnarray}
Here, $\varepsilon_{\vec k}$ is the normal state dispersion relation, and 
$\beta = 1/k_B T$. Given the form of $V_{\vec k,\vec k'}$, it is clear that 
the gap-function must be expandable in the basis functions for the irreducible
representations of $C_{4v}$ in the following way
\begin{eqnarray}
\Delta_{\vec k} & = & \sum_{\eta} ~ \Delta_{\eta} ~ B_{\eta}(\vec k),
\end{eqnarray}
where $\Delta_{\eta}$ are amplitudes that must be found selfconsistently. 
Inserting such an Ansatz back into the gap-equation and equating coefficients
of the linearly independent functions $\{ B_{\eta}(\vec k) \}$, one finds the 
following coupled, nonlinear algebraic equations for the gap-amplitudes 
$\Delta_{\eta}$
\begin{eqnarray}
\Delta_{\eta} & = & \sum_{\eta'} ~ \Delta_{\eta'} ~ M_{\eta \eta'} \nonumber \\
M_{\eta \eta'} & \equiv & - \lambda_{\eta} ~ \sum_{\vec k} ~ 
B_{\eta}(\vec k) ~ B_{\eta'}(\vec k) ~ \chi_{\vec k}.
\end{eqnarray}
Notice that the pair-susceptibility $\chi_{\vec k}$ transforms as an s-wave 
function. Hence, the matrix $M_{\eta \eta'}$ must be block-diagonal by symmetry, 
in the s-wave and d-wave sectors. Including the five basis functions we have 
used in this paper, the matrix thus block-diagonalizes into a $3 \times 3$ and 
a $2 \times 2$ matrix. 

Consider first the case  $T=T_c$, where the equations are linear.
The various sectors must be solved separately for the s-wave 
gap and the d-wave gap, respectively. The physically relevant solution is the 
one with the largest value of $T_c$, while the other solution is unphysical. 
The alternative is a scenario with two distinct superconducting transitions in 
zero magnetic field,  which is not acceptable. Therefore one can never get a 
mixing of s-wave and d-wave symmetry on a perfectly square lattice at
the critical temperature. The possible exception is the one where coupling 
constants in combination with filling fractions have been fine-tuned in such 
a way that the two sectors yield the same $T_c$. 

Below $T_c$, the situation is changed. The matrix $M_{\eta \eta'}$ is still
block diagonal, but there is nonetheless coupling between the s-wave and
d-wave sectors, since the pair-susceptibility depends on the entire
gap $\Delta_{\vec k} = \sum_{\eta} ~ \Delta_{\eta} ~ B_{\eta}(\vec k)$.
As a consequence,  at $T=T_c$ the dominant channel becomes superconducting, 
while more components are added to the gap, at distinct temperatures,  as the 
temperature is lowered. This means that one is getting superconducting 
condensation in progressively more channels. Nonetheless,
these subsequent additions 
of channels to the order parameter do not constitute separate superconducting 
critical points. They simply add more amplitude to $|\Delta_{\vec k}|$. 
In mathematical terms, this is equivalent to a bifurcation in the solution of 
the nonlinear gap-equation. Such bifurcations have recently been studied for
the BCS-gap equation \cite{Spathis:1992}. 

We have solved the coupled mean-field gap equations numerically, to obtain 
the superconducting gap, and hence various thermodynamic quantities. 
We omit details of our calculations of the gap-function $\Delta_{\vec k}$ 
itself, suffice it to say that we have reproduced in detail the results of 
Ref.  \cite{O'Donovan:1995}. The free energy, internal energy, entropy, 
specific heat, and critical magnetic fields are found by  using the standard
expressions for the free energy $F$ and the entropy $S$
\begin{eqnarray}
F&=&\sum_{\vec k}~(~\varepsilon_{\vec k}-\chi_{\vec k}~|\Delta_{\vec k}|^2 )
- \frac{1}{\beta} ~ \sum_{\vec k} ~ 
\ln[2 +\exp(\beta E_{\vec k}) +\exp(-\beta E_{\vec k}) ] \nonumber \\
S & = & - 2 k_B ~ \sum_{\vec k} ~ \bigl[ ~ (1-f_{\vec k}) ~ \ln (1-f_{\vec k})
+ f_{\vec k} ~ \ln f_{\vec k} \bigr], 
\end{eqnarray}
with $F=U-TS$, and $C = T dS/dT$, where $U$ is the internal energy,  and 
$f_{\vec k} = 1/[1+\exp(\beta E_{\vec k})]$ is the momentum distribution 
function for the particle-like elementary excitations of the superconductor.

The lower  and upper critical fields $B_{c1}$ and $B_{c2}$, respectively, are 
found from the standard expressions $F(T=T_c) - F(T<T_c) = B_{c1}^2/2 \mu_0$, 
and $B_{c2} = \Phi_0/2 \pi \xi^2$, where we have calculated the superconducting
coherence length from $\xi= {\rm v}_F/\Delta$. Here ${\rm v}_F$ is the 
Fermi-velocity in the normal metallic state, while $\Delta$ is taken to be the 
absolute value of the gap $\Delta_{\vec k}$ averaged over the Fermi-surface. As 
a sideproduct, we obtain the Ginzburg-Landau ratio $\kappa = B_{c2}/\sqrt{2} 
B_{c1}$.

Our results exhibiting the quantities described above, are shown in 
Figs. \ref{fig:thermoU0}-\ref{fig:kappaU4}. 
As the temperature is lowered, one will observe cusps in $|\Delta_{\vec k}|$, 
and hence the internal energy and entropy, at the temperatures where new 
channels are coupled into the gap. The specific heat will show 
BCS-discontinuities at all the temperatures where the new channels 
condense, in addition to the BCS-discontinuity at the superconducting 
transition \cite{Spathis:1992}. However, the amplitude-fluctuations of the 
orderparameter are massive at these lower temperatures, and the superconducting 
correlation length is finite. This is a generic feature of superconductivity 
in systems with competing pairing channels, irrespective of whether the 
dominant pairing is d-wave or s-wave. 

\section{Specific heat}
For the case of extreme type-II superconductors, critical fluctuations will 
surely modify the results for the specific heat close to the critical 
temperature. The anomaly in the specific heat at the critical temperature is 
therefore not expected to be of the BCS-type at all, due to phase-fluctuations 
of the order parameter. As recently emphasized \cite{Kivelson:1997}, in extreme 
type-II superconductors with a low superfluid density, the dominant 
contributions to the specific heat anomaly are expected to be phase-fluctuations 
in the order parameter, not amplitude-fluctuations as in BCS. 
This is easily seen by noting that the superfluid stiffness is related to the
free energy $F$ and a phase-twist $\delta \phi$ in the order parameter 
across the system via the Fisher-Barber-Jasnow relation
\cite{Fisher:1973}
\begin{eqnarray}
\rho_s = \Biggl( \frac{\partial^2 F}
                  {\partial (\delta \phi)^2} \Biggr)_{\delta \phi = 0}.
\end{eqnarray}
On the other hand, superconductors arising from poor conductors such as the 
high-$T_c$ compounds, do in fact have a low superfluid stiffness 
\cite{Kivelson:1997}. This implies that phase-fluctuations in the 
superconducting orderparameter are soft, dominating the fluctuation 
spectrum, while amplitude fluctuations may be neglected. One may think of 
the superconductivity as arising out of a quantum fluid of preformed pairs.
Hence, for weakly coupled layers such as those we have considered in this 
paper, the critical point is in the universality class of the $3D$ XY-model, 
and the specific heat should have an analagous critical anomaly, which turns 
out to be  {\it an asymmetric  logarithmic singularity} with a specific 
heat exponent $\alpha = - 0.007(6)$ rather than the finite discontinuity 
of the BCS-type \cite{LeGillou:1980,Nguyen:1997}. This has recently also 
been observed in  experiments on the extreme type-II $Y Ba_2 Cu_3 O_7$ 
cuprate  with a Ginzburg-Landau parameter $\kappa \approx 50$ 
\cite{Roulin:1997}. (The rather counter-intuitive result that 
{\it fluctuations} may convert and {\it sharpen} a mean-field like 
{\it finite discontinuity} in the specific heat into a (logarithmic) 
{\it singularity} at the transition, rather 
than smoothing over the asperities in the specific heat, is known from 
Onsager's famous solution of the $2D$ Ising model \cite{Onsager:1944}.)

The weak anomalies at the lower temperatures should however be well captured 
by the mean-field theory, provided they are located outside the critical regime, 
which is likely to be the case. This is therefore a way of distinguishing 
intrinsic anomalies in the specific heat from those anomalies one will observe 
in multi-phase compounds with a distribution of critical temperatures. In the 
former case, one should observe {\it one} $3D$ XY-like anomaly at the critical 
temperature, and additional weaker mean-field BCS-like anomalies at lower 
temperatures. In the latter case, one should observe a number of anomalies 
roughly of the same order of magnitude, all of the $3D$ XY-type.  

A large onsite repulsion $U$ ultimately suppresses pairing in the isotropic and 
extended s-wave channels. This may explain why two anomalies in the specific 
heat so far have not been observed  in zero magnetic field in $YBa_2Cu_3O_7$ 
\cite{Roulin:1997}. However one should note that an inclusion of the $W$-term 
in Eq. \ref{Hamiltonian} will give pairing in another d-wave channel, 
$d_{xy}$, which could compete with $d_{x^2-y^2}$ in the presence of a very 
large Hubbard-$U$, in the relevant doping regime, and hence also give two 
anomalies in the specific heat even in large-$U$ compounds such as the 
high-$T_c$ cuprates. Recent experiments show intriguing features in 
microwave conductivity as well as London penetration depth well inside the 
superconducting phase, which may be consistent with the above picture 
\cite{Srikanth:1996}. 

\section{Acknowledgements}
Support from the Research Council of Norway (Norges Forskningsr{\aa}d) 
Grants No. 110566/410 and 110569/410, is gratefully acknowledged. 
The authors thank G. Angilella for  discussions.

\medskip
\begin{center}
\bf{LIST OF FIGURES}
\end{center}

\begin{figure} 
\centering
\caption{The probability density $\omega(i,j)=|\psi(i,j)|^2$ 
of the Cooper-pair wave function with $U=0$ at different filling fractions. 
In Figs. (a) and (b) the wave function exhibits s-wave symmetry for $n=0.06$ 
and $n=0.12$, respectively. Figs. (c) - (f) all exhibit d-wave symmetry of the
wave function where $n=0.14$ in (c), $n=0.17$ in (d), $n=0.51$ in (e), and 
$n=0.85$ in (f).}   
\label{fig:U0muchange}
\end{figure}

\begin{figure} 
\centering
\caption{The binding energy of the Cooper-pair as a function 
of doping, at $U=0$. For $n<0.11$ the binding energy of the s-wave pairing 
is energetically favorable compared to d-wave pairing. As doping increases,
the d-wave pairing becomes favorable.}
\label{fig:enrgU0}
\end{figure}

\begin{figure} 
\centering
\caption{The binding energy of the Cooper-pair as a
function of doping, at $U=1.0*t$. For $n<0.11$ the binding energy of the s-wave
pairing is energetically favorable to d-wave pairing. As doping increases,
the d-wave pairing becomes favorable.}\label{fig:enrgU1}
\end{figure}

\begin{figure} 
\centering
\caption{The probability density $\omega(i,j) = |\psi(i,j)|^2$ 
of the Cooper-pair wave function with $U=4.0*t$ at different filling fractions. 
Figs. (a) - (f) all exhibit d-wave symmetry of the wave function where 
$n=0.06$ in (a), $n=0.12$ in (b), $n=0.14$ in (c), $n=0.17$ in (d), 
$n=0.51$ in (e), and $n=0.85$ in (f).}   
\label{fig:U4muchange}
\end{figure}

\begin{figure} 
\centering
\caption{The probability density $\omega(i,j)=|\psi(i,j)|^2$ 
of the Cooper-pair wave function with $n=0.06$ at different onsite repulsions. 
For $U=0$, s-wave pairing is favorable (a). For the values
$U=0.50*t$,$U=0.70*t$,$U=1.0*t$ and $U=1.30*t$ as shown in (b), (c),
(d) and (e),  respectively, the
Figs. exhibit s-wave symmetry. For $U=1.45*t$ the Cooper pair wave
function tranforms as a d-wave.}
\label{fig:mu002changeU}
\end{figure}

\begin{figure} 
\centering
\caption{The binding energy of the Cooper-pair with
$n=0.06$ as a function of onsite repulsion. Since the onsite potential
couples to an s-wave, the d-wave binding energy is not affected by 
increasing $U$. For $U<1.4*t$, s-wave paring is favored. For larger
$U$'s, d-wave pairing is favored.}
\label{fig:enrgmu002}
\end{figure}

\begin{figure} 
\centering
\caption{The projected densities of states $N_{22}$ and
$N_{44}$ along with the single particle density of states $N(\varepsilon) 
= \sum_{\vec k} \delta(\varepsilon - \varepsilon_{\vec k})$.}
\label{fig:ProDos}
\end{figure}

\begin{figure} 
\centering
\caption{Thermodynamic quantities in superconducting and
normal states for $n=0.25$ and $U=0$. Figs. (a)- (d) show the entropy, free
energy, internal energy and specific heat, respectively. They all
exhibit a critical behavior at $T\approx90~K$. In (d) it can easily
be seen that another symmetry channel switches on at $T\approx60~K$, giving 
rise to cusps in the internal energy and entropy, and a discontinuity in the 
specific heat. Figs. (e) and (f) show the coherence length and the critical 
fields, respectively. Note that the coherence length has singular behavior 
only at $T \approx 90K$.}
\label{fig:thermoU0}
\end{figure}

\begin{figure} 
\centering
\caption{Thermodynamic quantities in superconducting and
normal states for $n=0.60$ and $U=4.0*t$. Figs. (a)- (d) show the entropy, free
energy, internal energy and specific heat, respectively. They all
exhibit a critical temperature at $T\approx90~K$. Figs. (e) and (f)
show the coherence length and the critical fields, respectively. In this case, 
there are no additional features below $T_c$, due to the complete 
suppression of the instability in the competing s-wave channel as a result 
of the large onsite $U$.}
\label{fig:thermoU4}
\end{figure}

\begin{figure} 
\centering
\caption{The Ginzburg-Landau parameter, $\kappa$,
as a function of temperature for the cases: (a) $n=0.25$ and $U=0$ and
(b) $n=0.60$ and $U=4.0*t$. The result shows that the superconductor model 
is of the extreme type-II variety. Hence, one expects the dominant critical 
fluctuations in the order parameter to be phase-fluctuations, not amplitude 
fluctuations.}
\label{fig:kappaU4}
\end{figure}

%\bibliographystyle{prsty}
%\bibliography{bibnames}

\include{Allfigs}

\end{document}

%% file: Allfigs.tex
\clearpage

\setlength{\textwidth}{17cm}
\pagestyle{empty}
\setcounter{figure}{0}

\begin{figure}
\enlargethispage{2cm}
\centering
\subfigure[]{\epsfig{figure=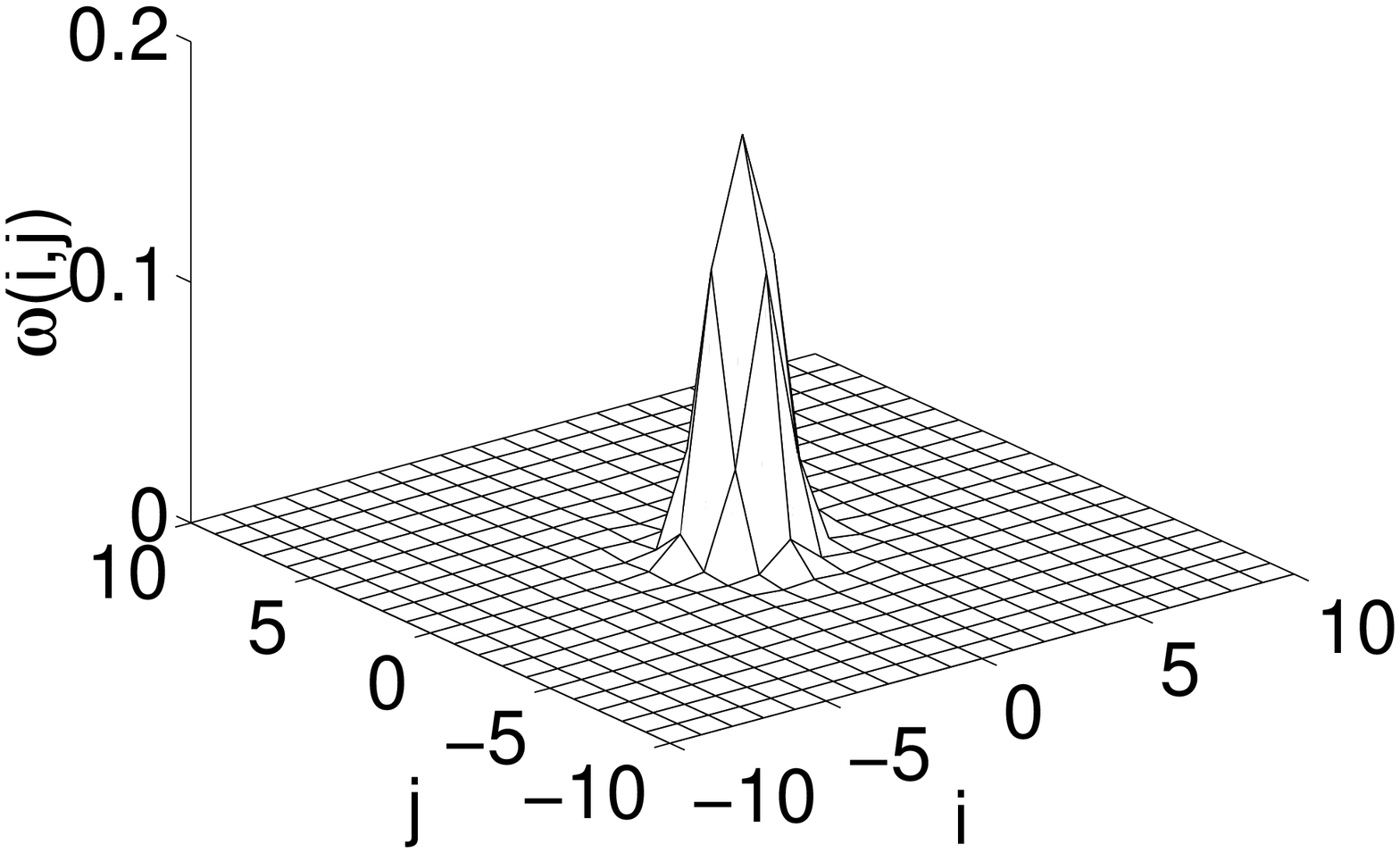,width=0.46\textwidth}}\quad
\subfigure[]{\epsfig{figure=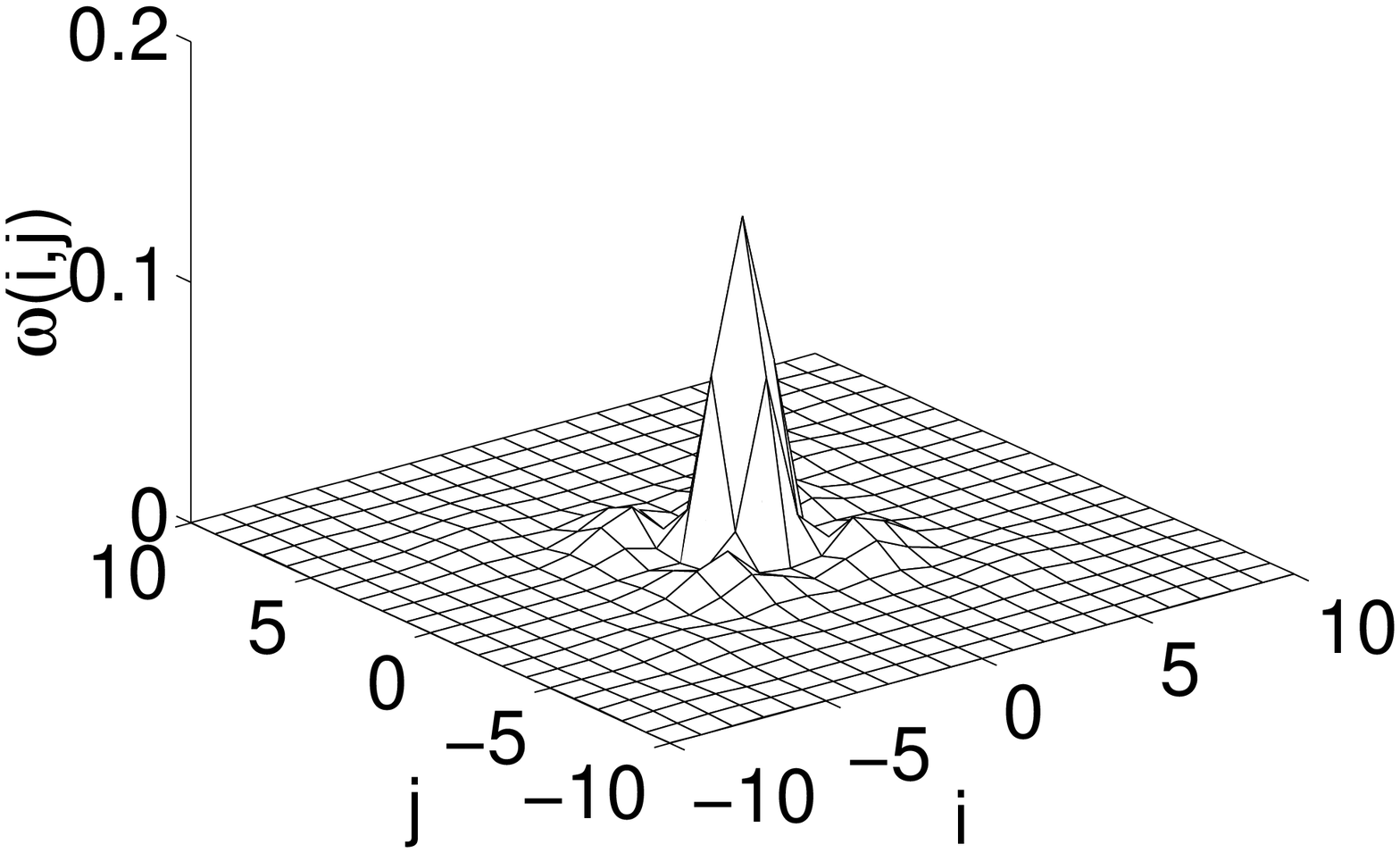,width=0.46\textwidth}}\\
\subfigure[]{\epsfig{figure=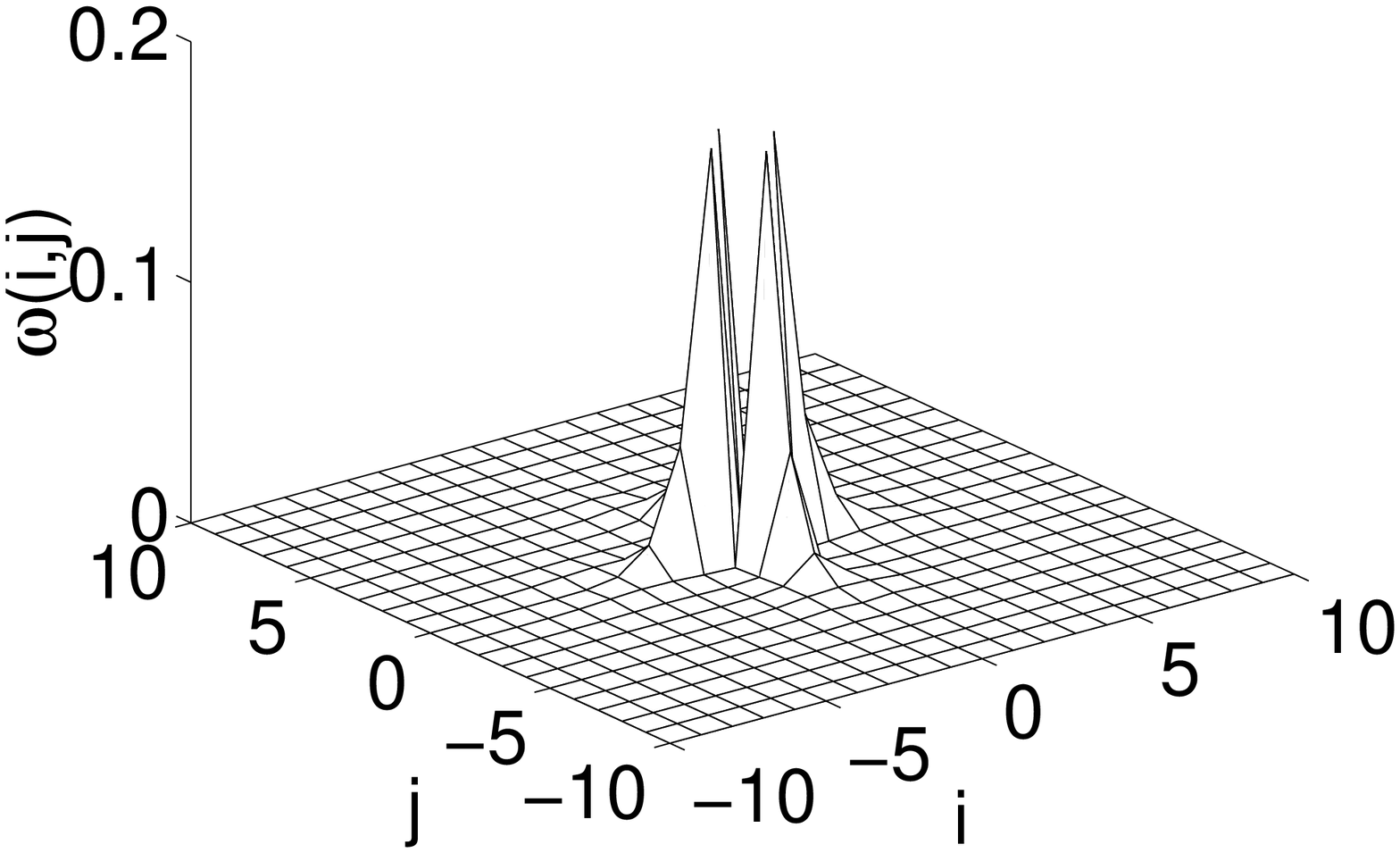,width=0.46\textwidth}}\quad
\subfigure[]{\epsfig{figure=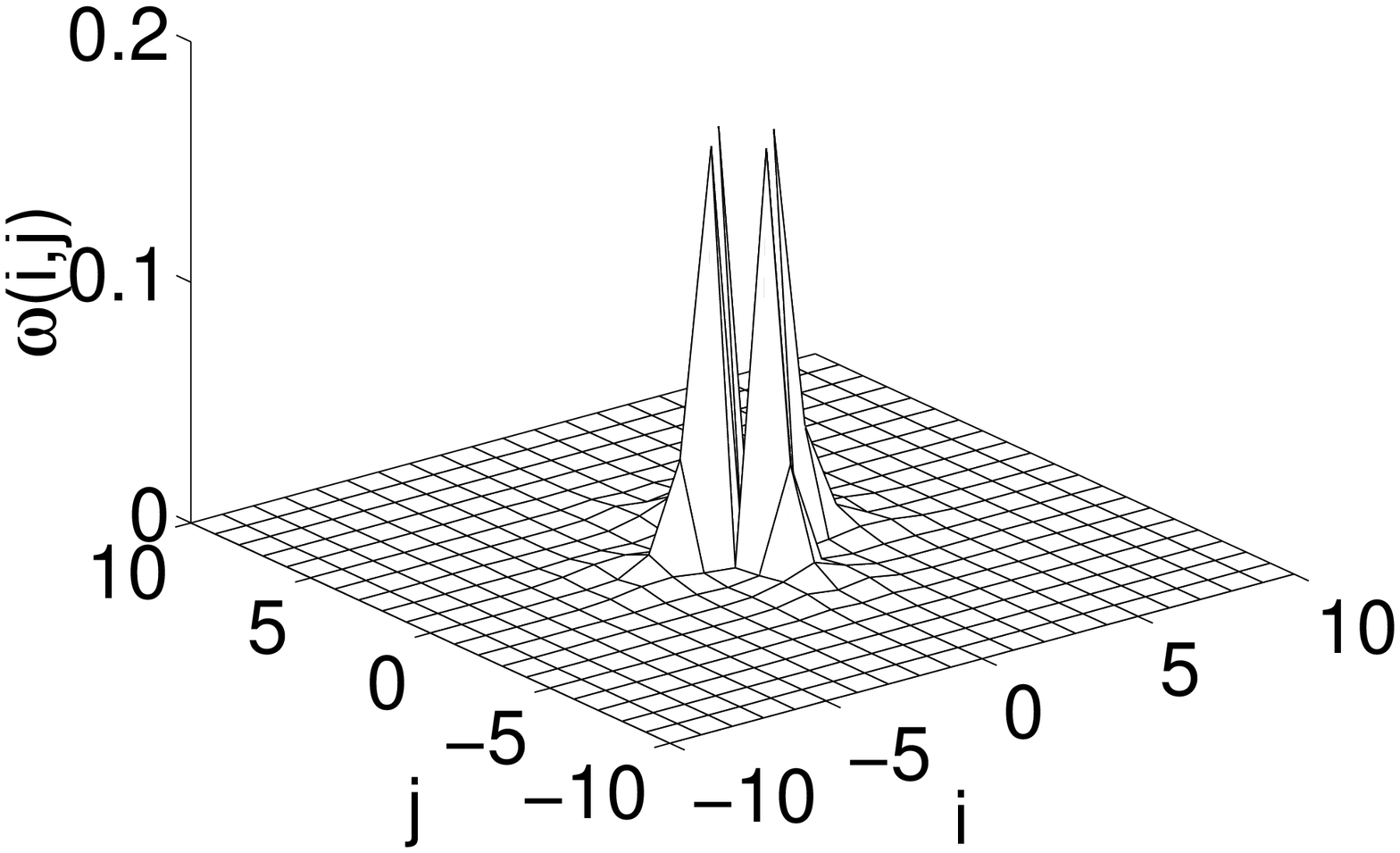,width=0.46\textwidth}}\\
\subfigure[]{\epsfig{figure=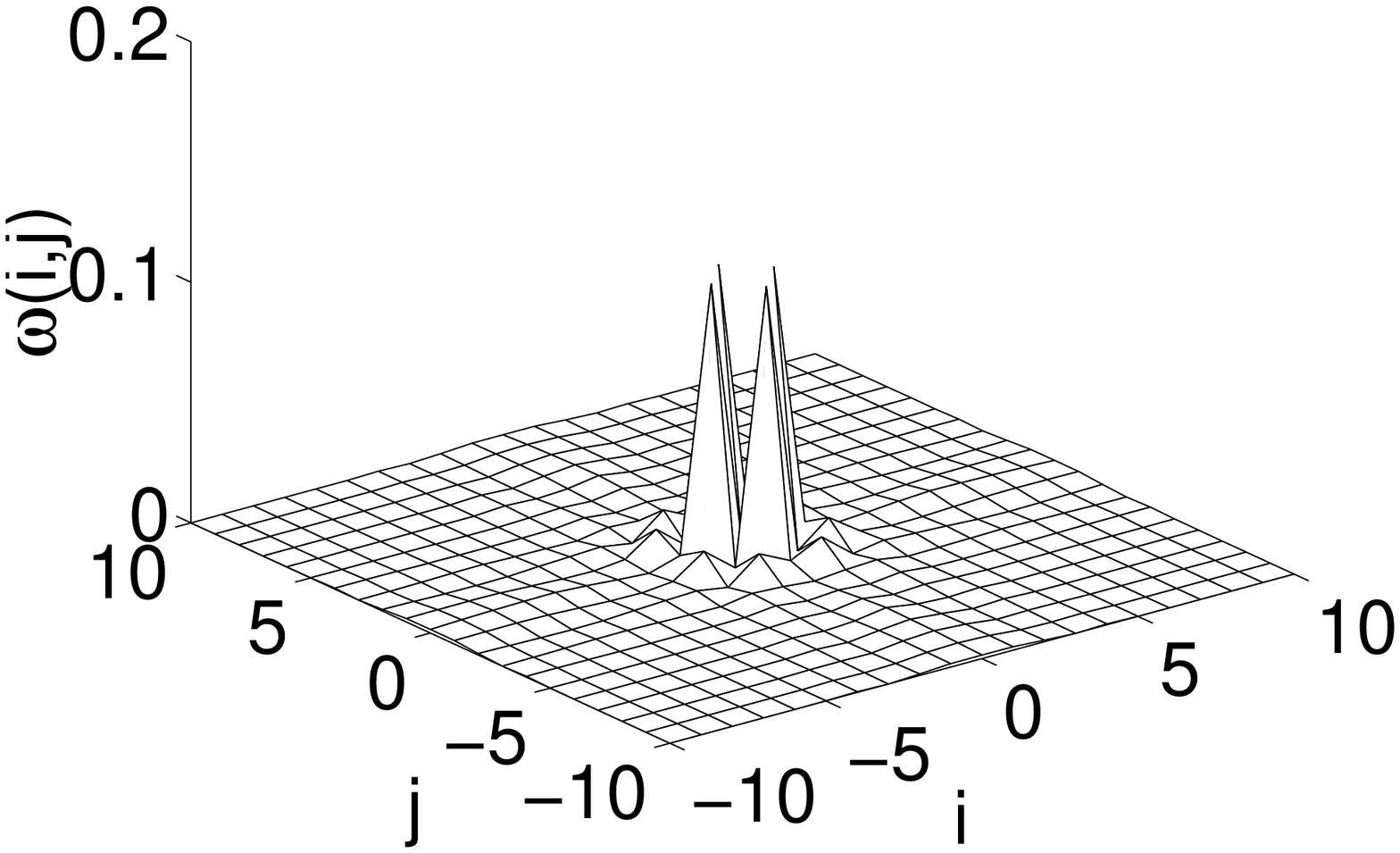,width=0.46\textwidth}}\quad
\subfigure[]{\epsfig{figure=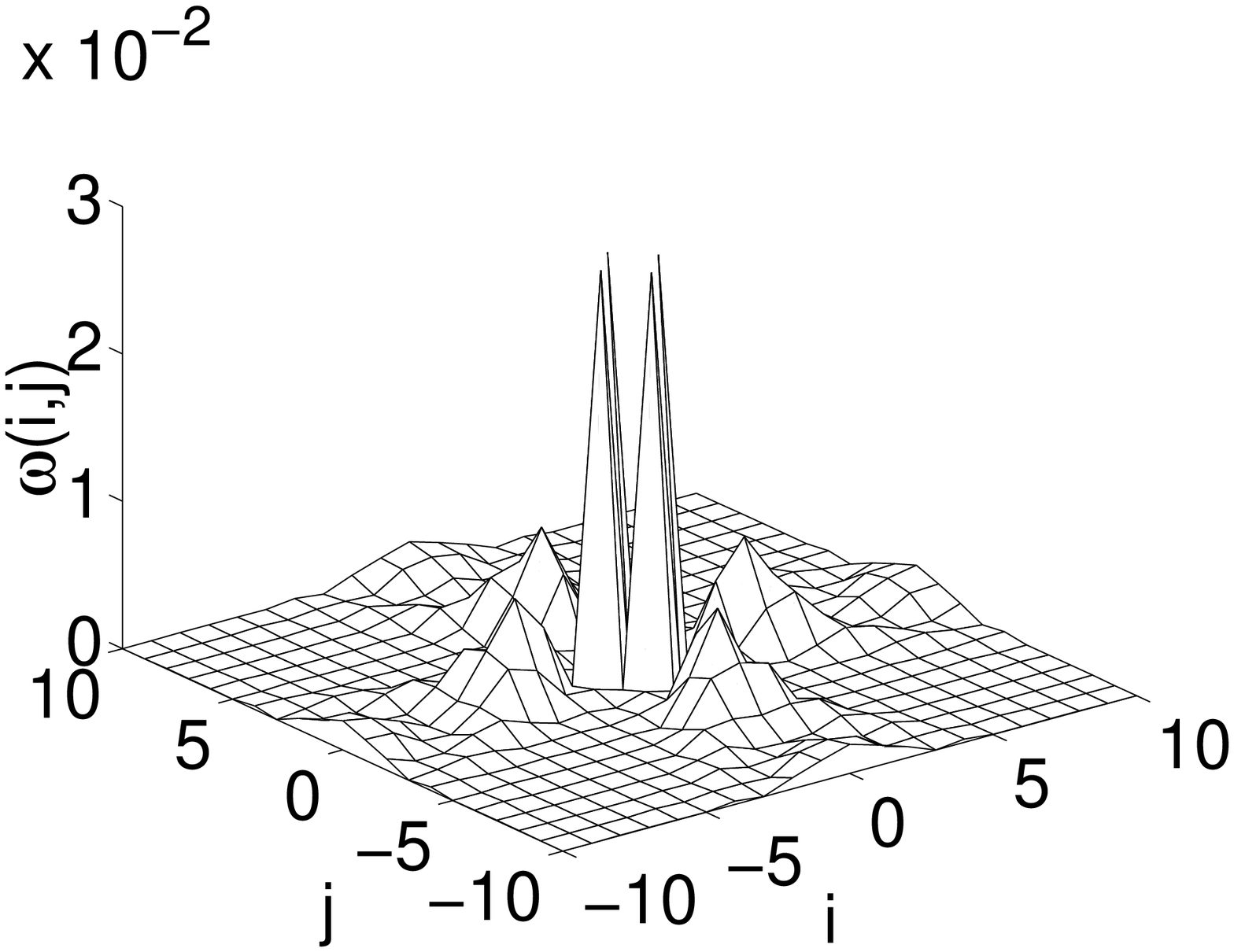,width=0.46\textwidth}}\\
\caption{}
\end{figure}
\clearpage

\begin{figure}
\enlargethispage{2cm}
\begin{center}
\epsfig{figure=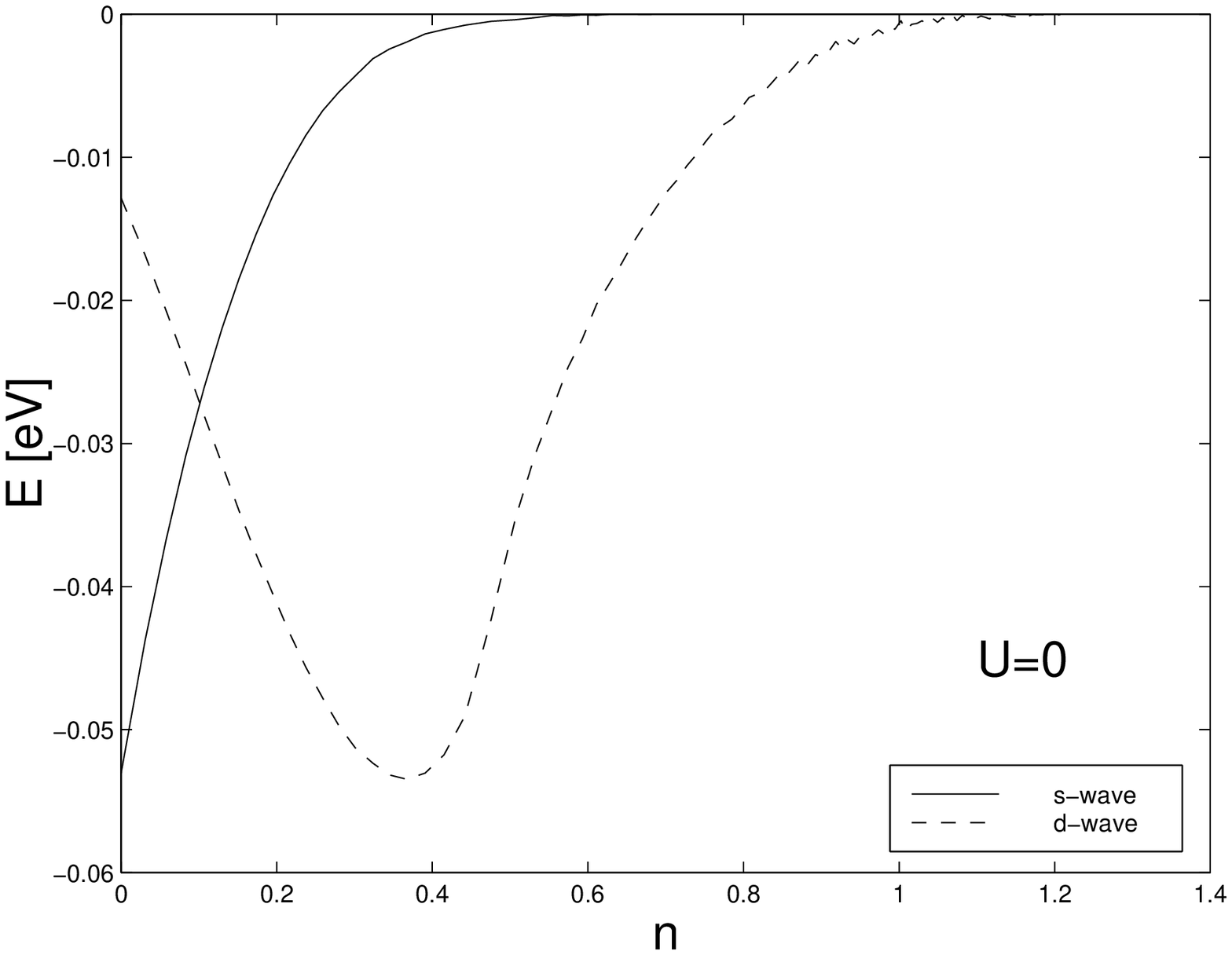,width=0.7\textwidth}
\end{center}
\caption{}
\end{figure}

\begin{figure}
\centering
\epsfig{figure=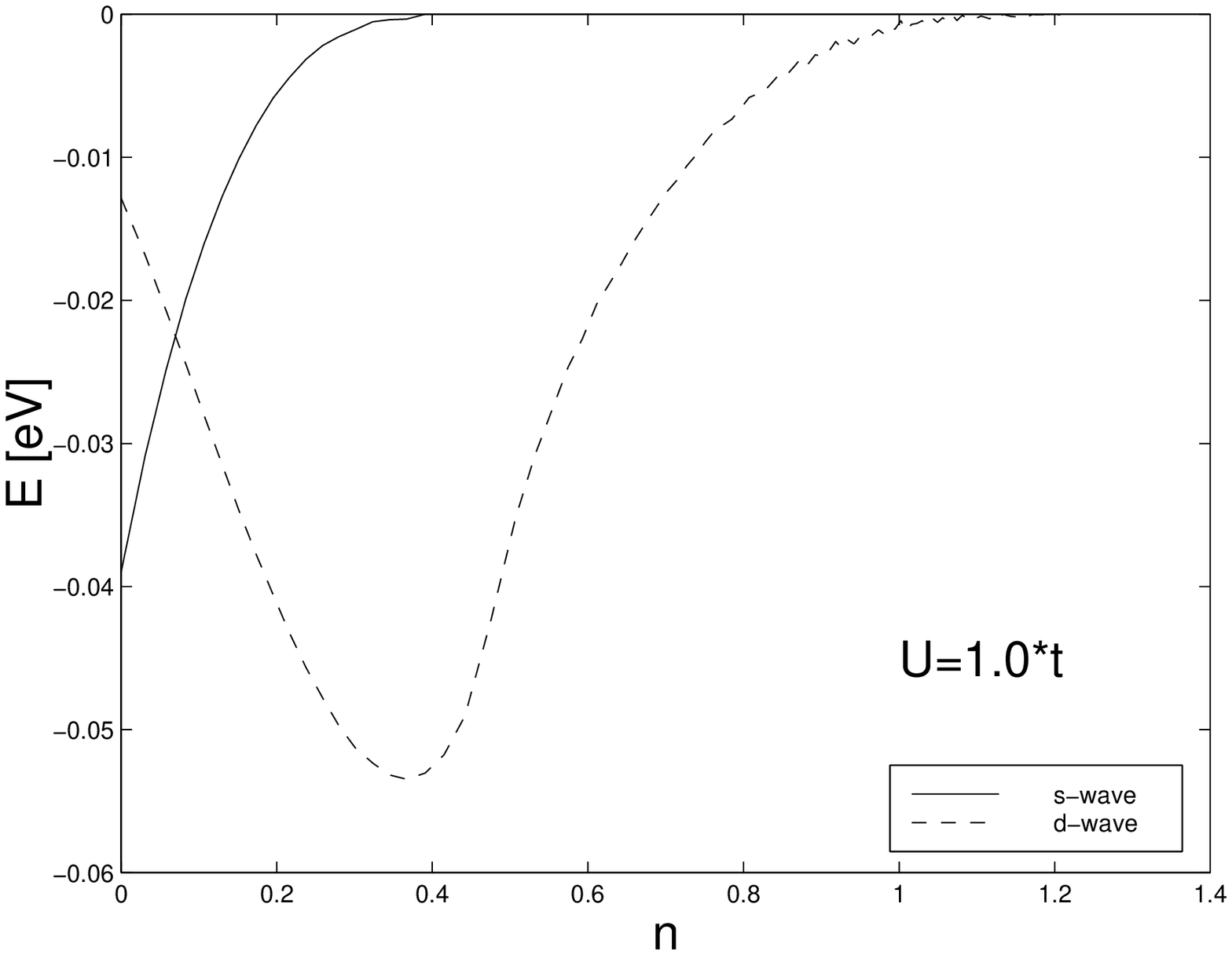,width=0.7\textwidth}
\caption{}
\end{figure}
\clearpage

\begin{figure}
\enlargethispage{2cm}
\centering
\subfigure[]{\epsfig{figure=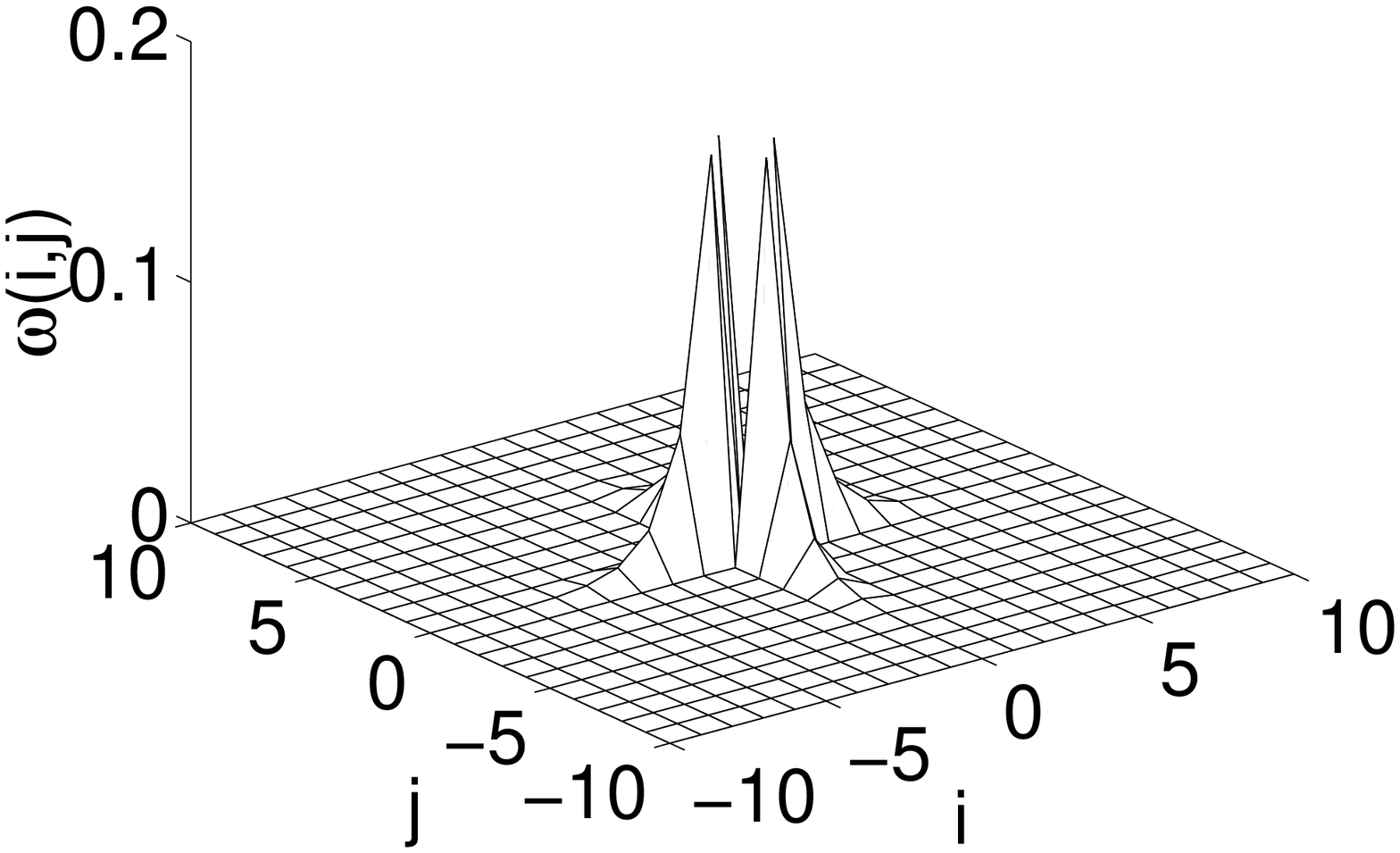,width=0.46\textwidth}}\quad
\subfigure[]{\epsfig{figure=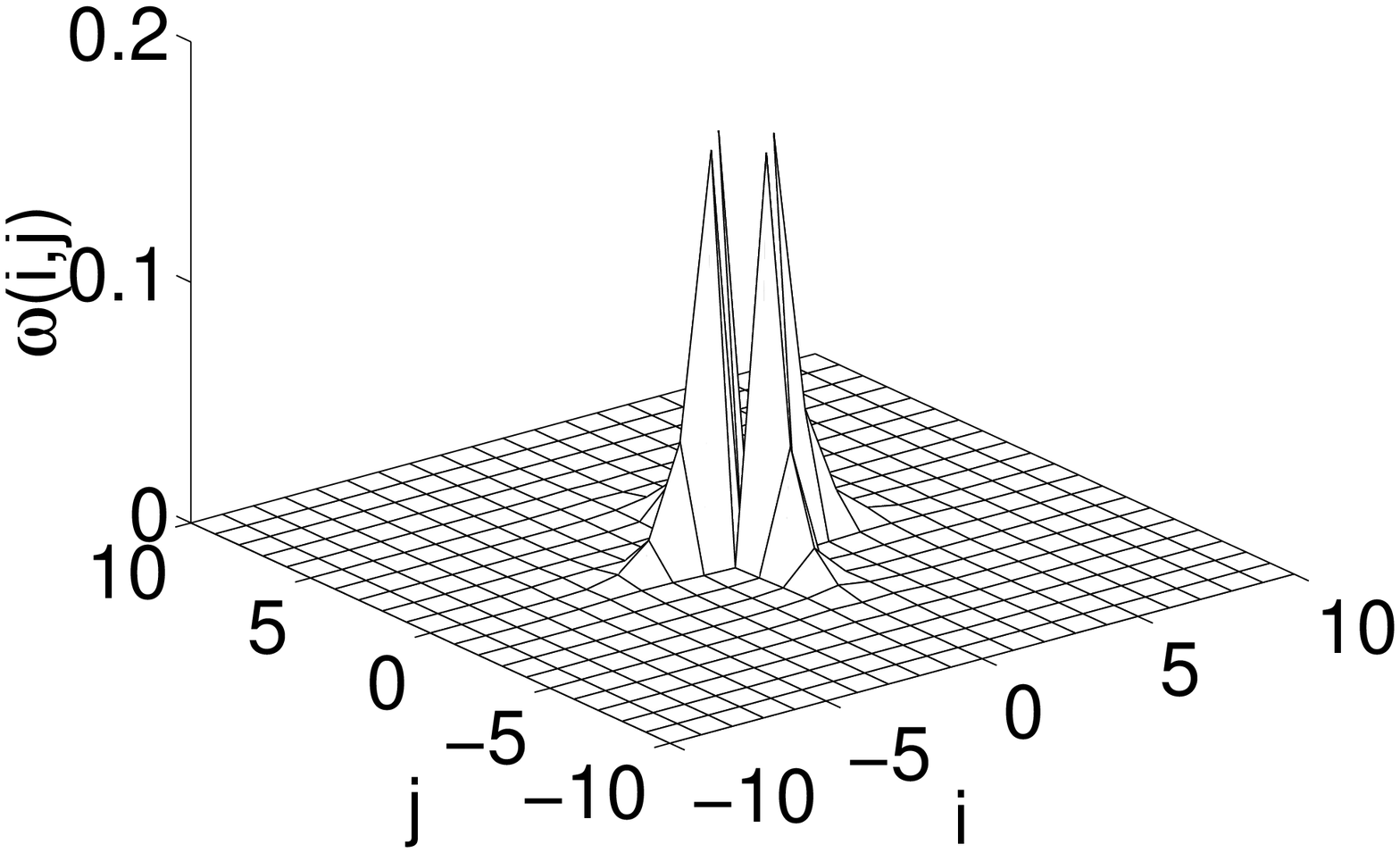,width=0.46\textwidth}}\\
\subfigure[]{\epsfig{figure=cwU4mu005.eps,width=0.46\textwidth}}\quad
\subfigure[]{\epsfig{figure=cwU4mu007.eps,width=0.46\textwidth}}\\
\subfigure[]{\epsfig{figure=cwU4mu021.eps,width=0.46\textwidth}}\quad
\subfigure[]{\epsfig{figure=cwU4mu045.eps,width=0.46\textwidth}}\\
\caption{}
\end{figure}
\pagebreak

\begin{figure}
\enlargethispage{2cm}
\centering
\subfigure[]{\epsfig{figure=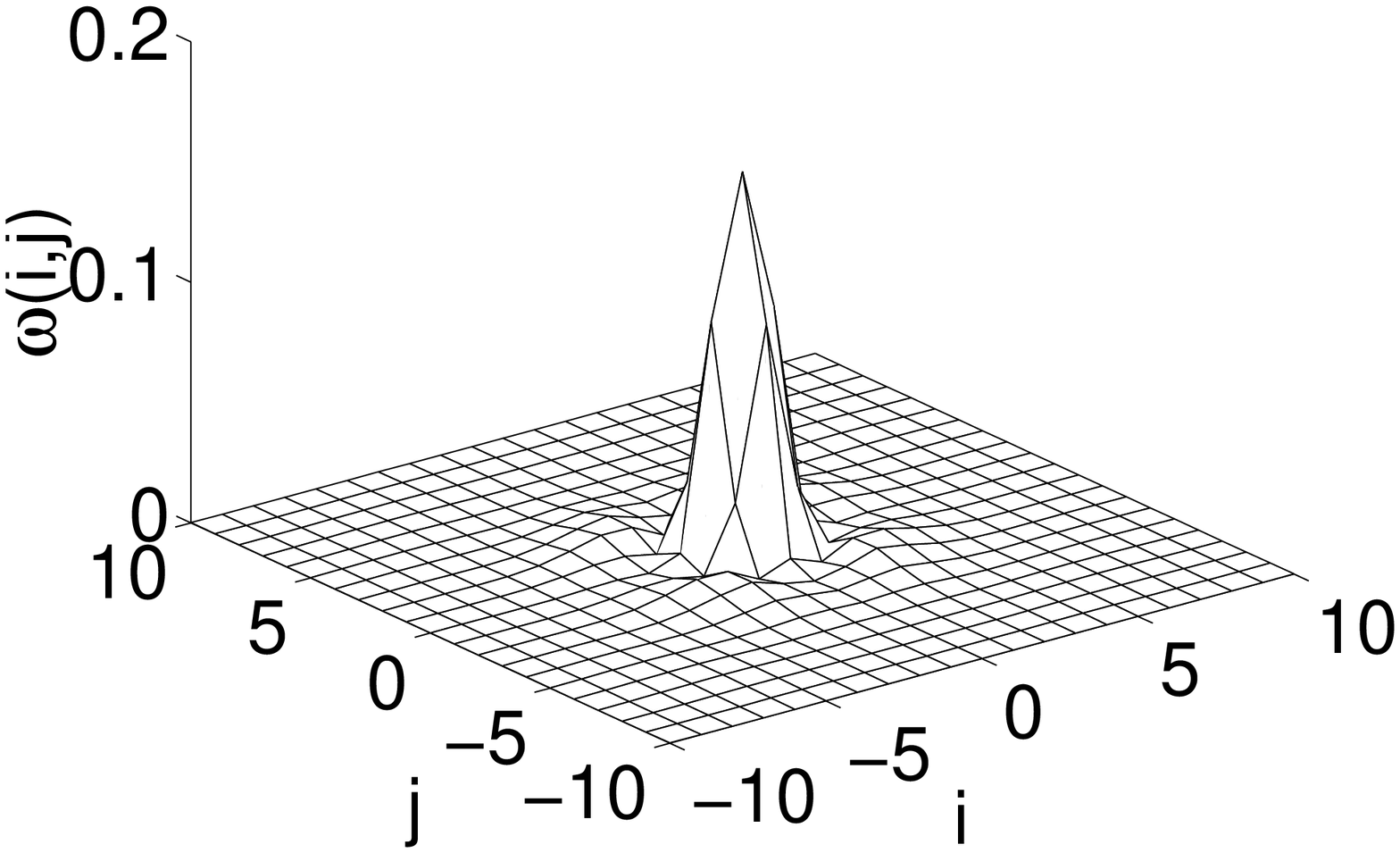,width=0.46\textwidth}}\quad
\subfigure[]{\epsfig{figure=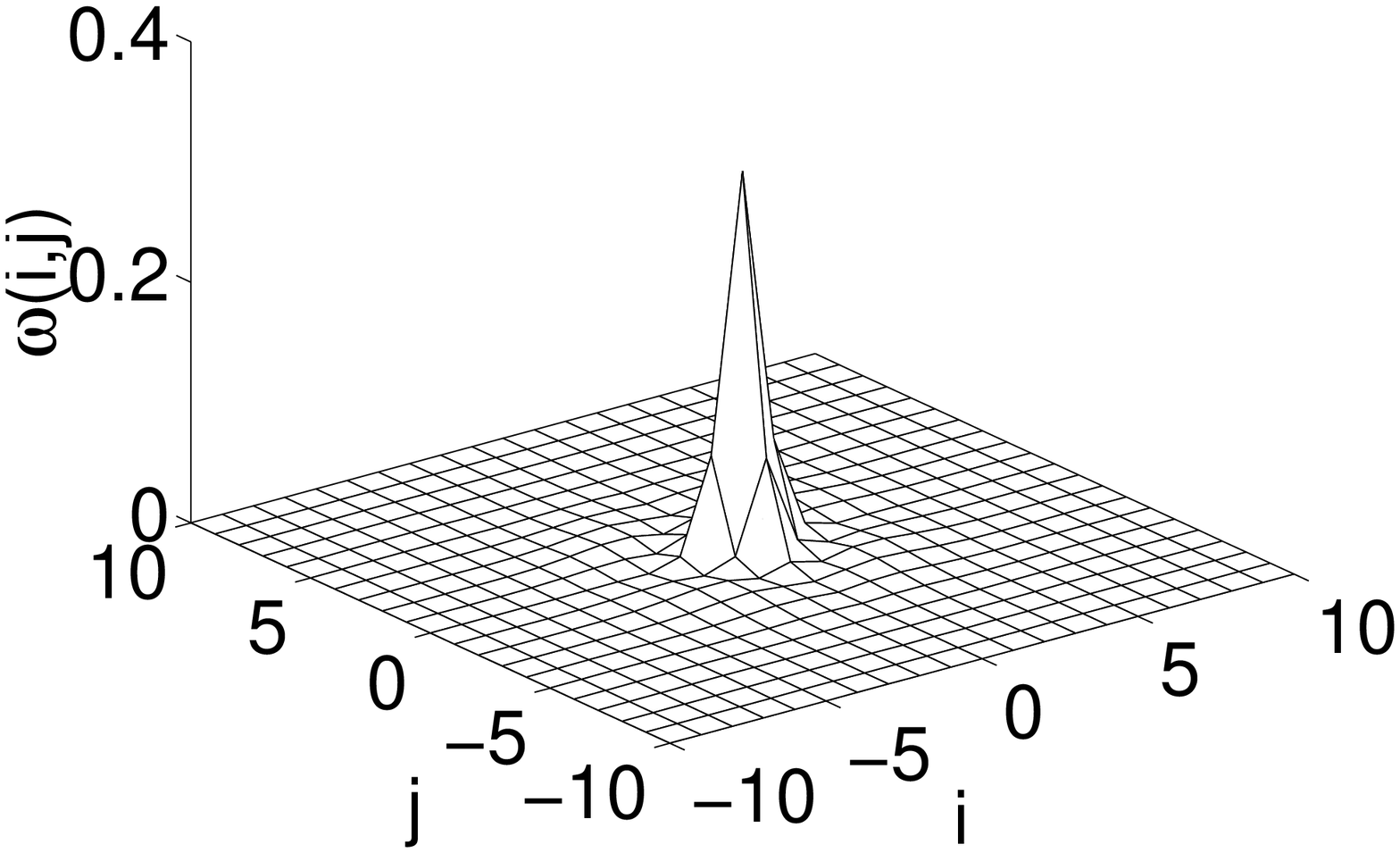,width=0.46\textwidth}}\\
\subfigure[]{\epsfig{figure=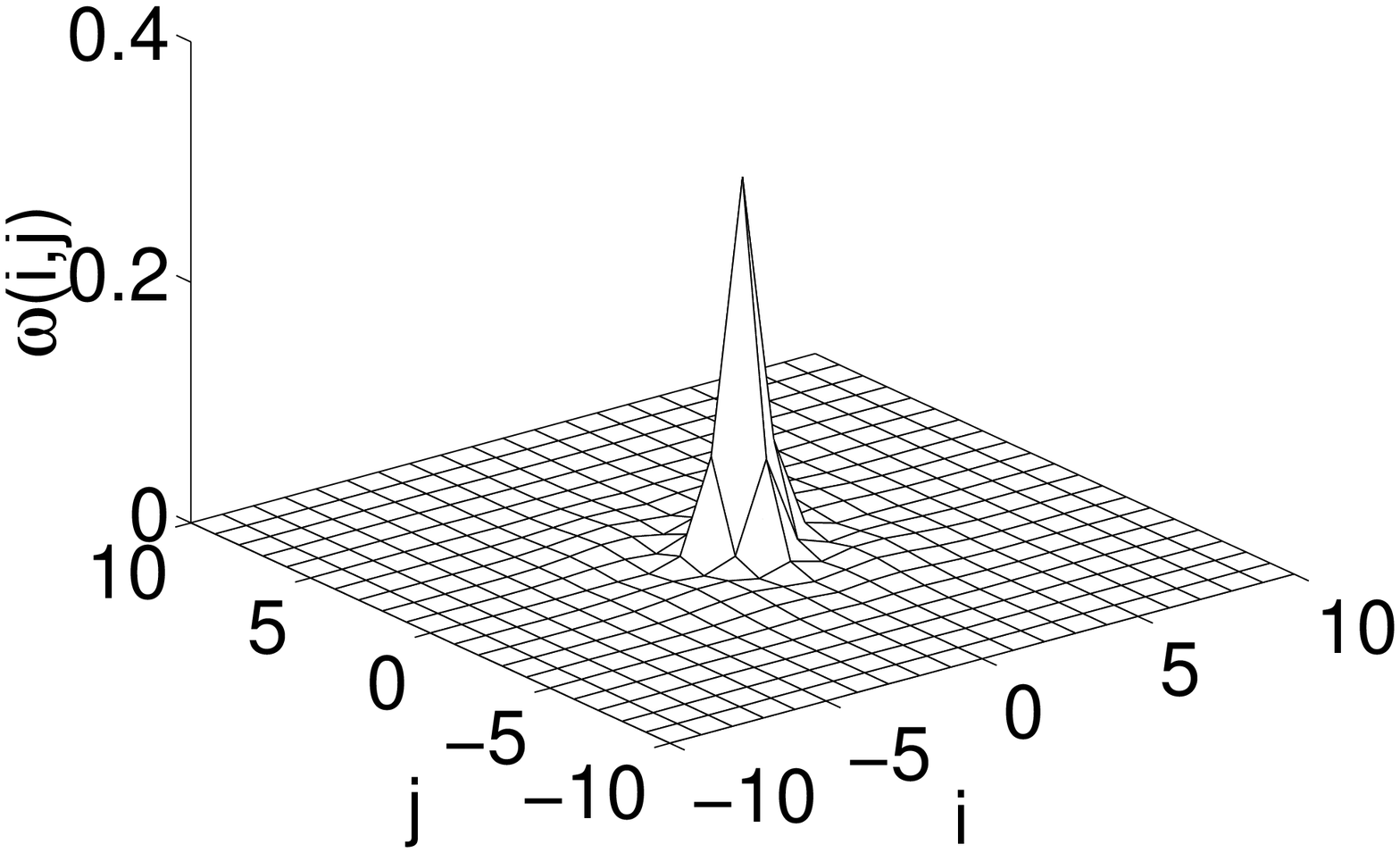,width=0.46\textwidth}}\quad
\subfigure[]{\epsfig{figure=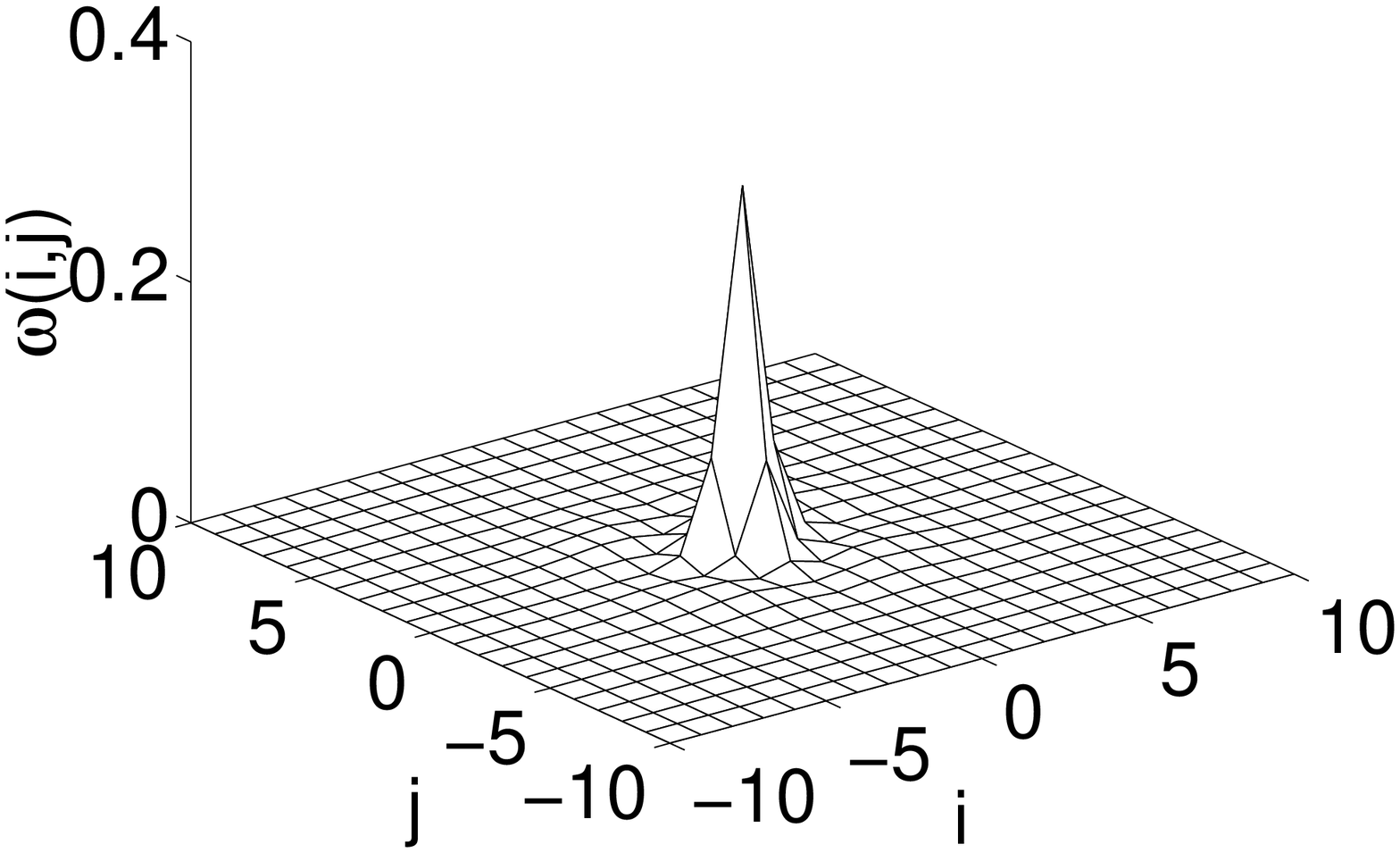,width=0.46\textwidth}}\\
\subfigure[]{\epsfig{figure=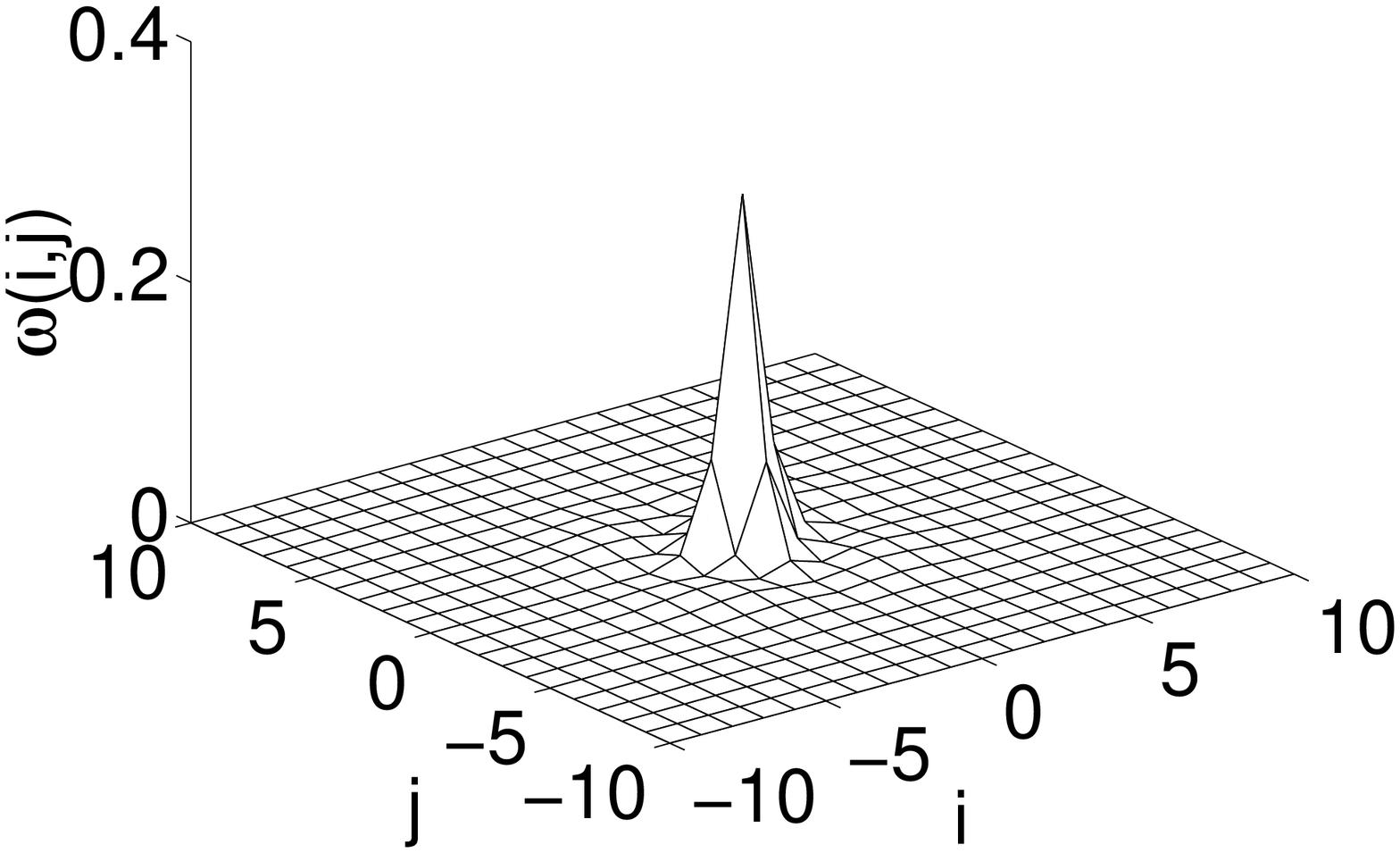,width=0.46\textwidth}}\quad
\subfigure[]{\epsfig{figure=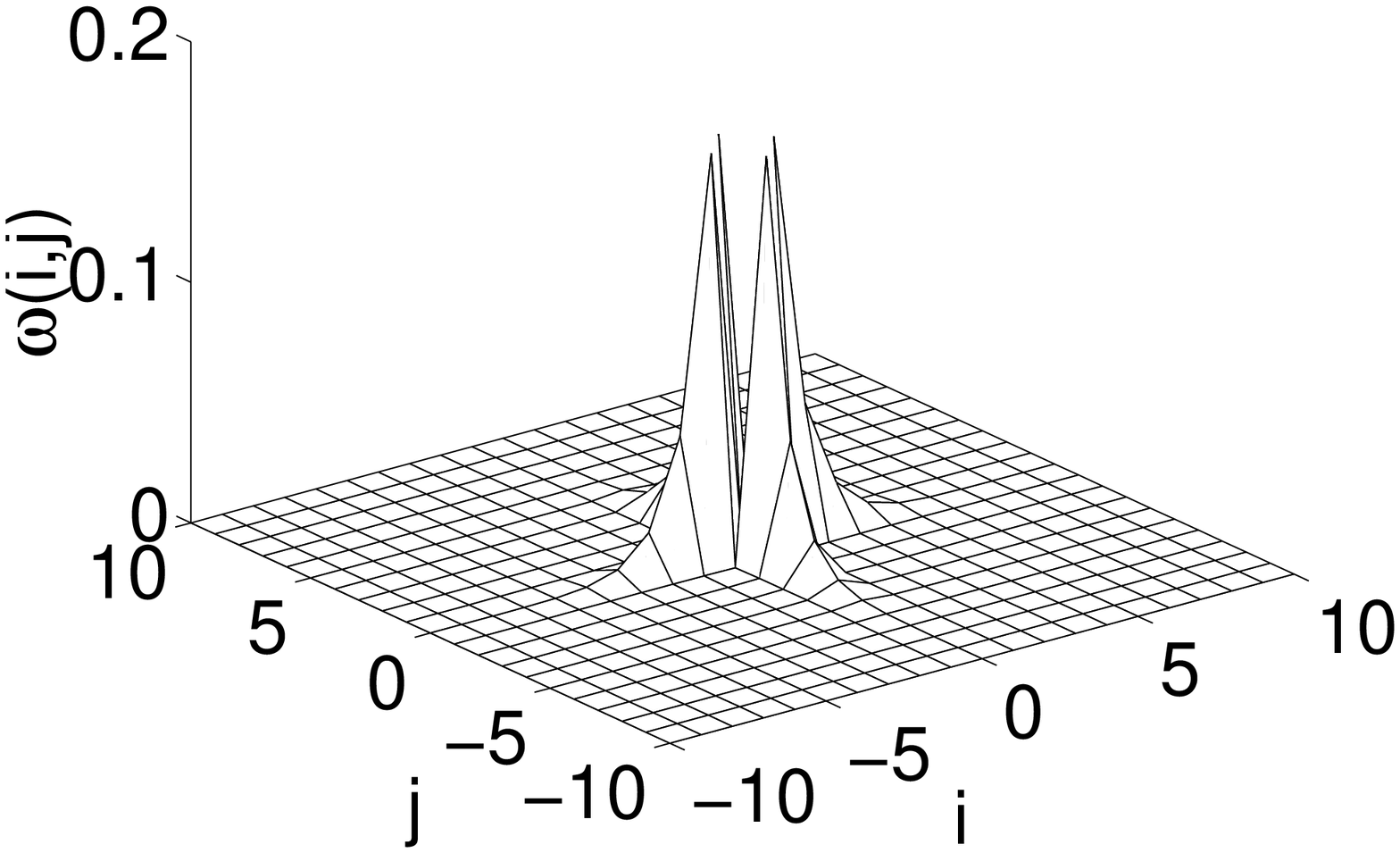,width=0.46\textwidth}}\\
\caption{}
\end{figure}
\clearpage

\begin{figure}
\enlargethispage{3cm}
\centering
\epsfig{figure=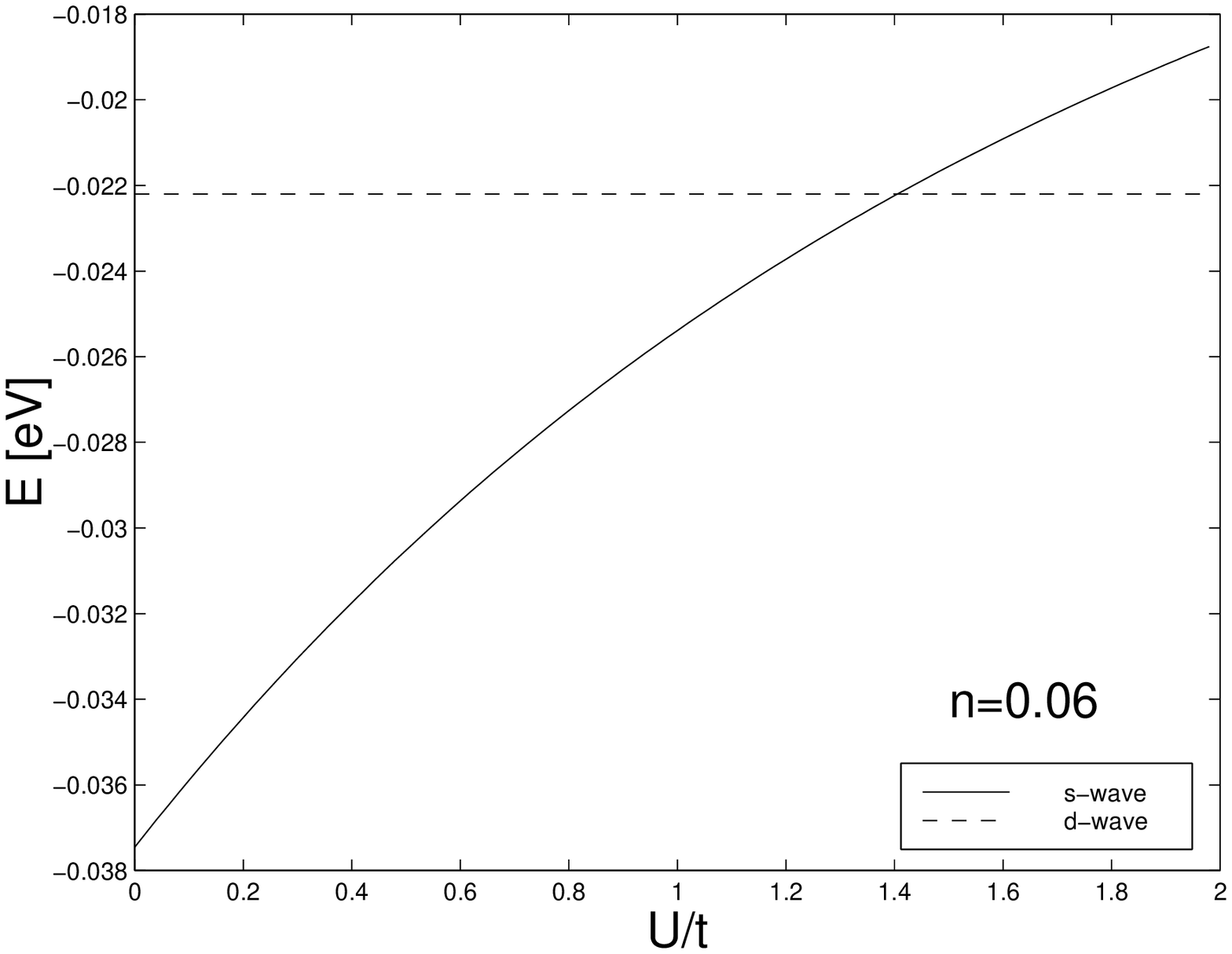,width=0.7\textwidth}
\caption{}
\end{figure}

\begin{figure}
\centering
\epsfig{figure=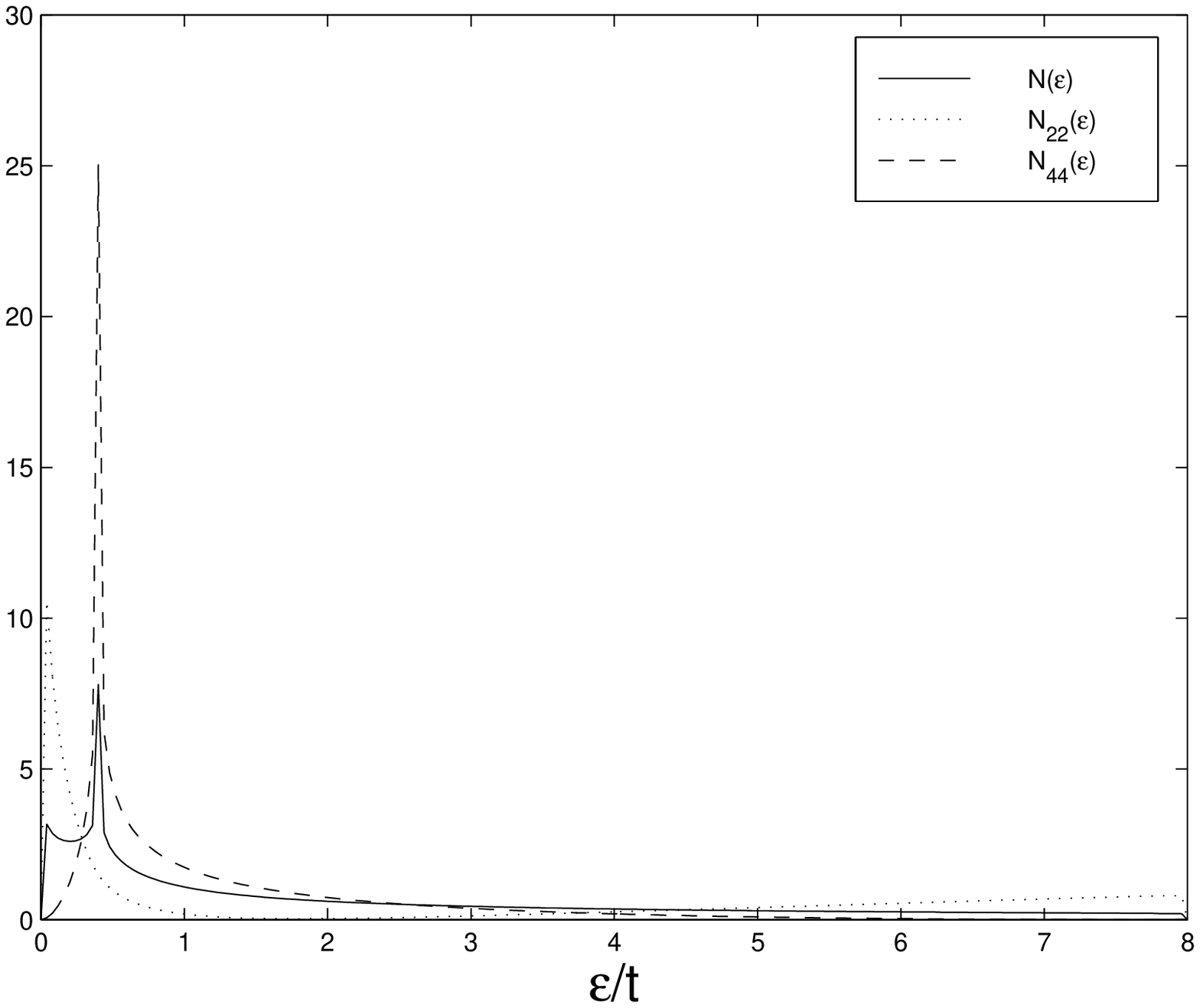,width=0.7\textwidth}
\caption{}
\end{figure}

\clearpage
\enlargethispage{3cm}
\begin{figure}
\centering
\subfigure[]{\epsfig{figure=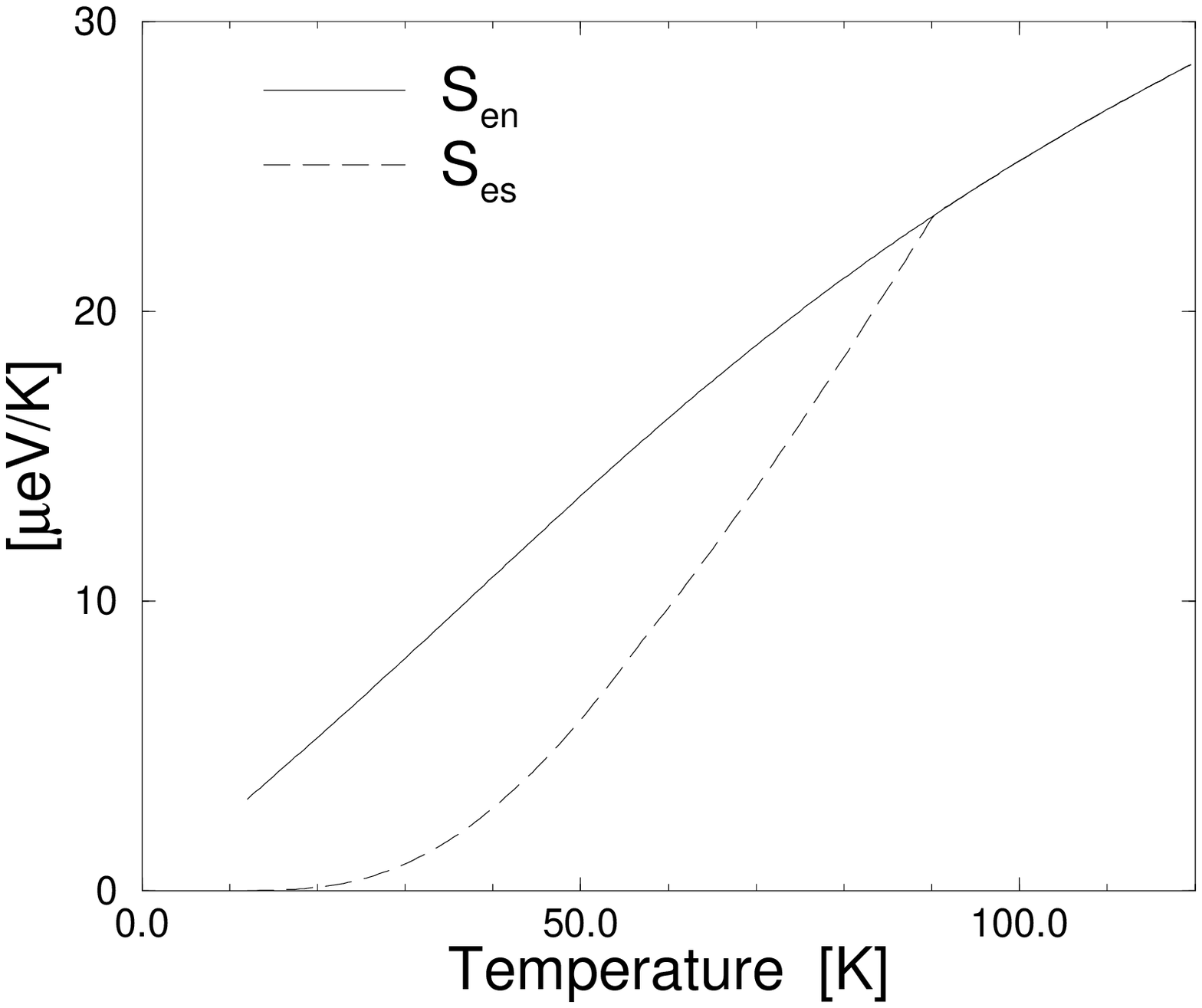,width=0.46\textwidth}}\quad
\subfigure[]{\epsfig{figure=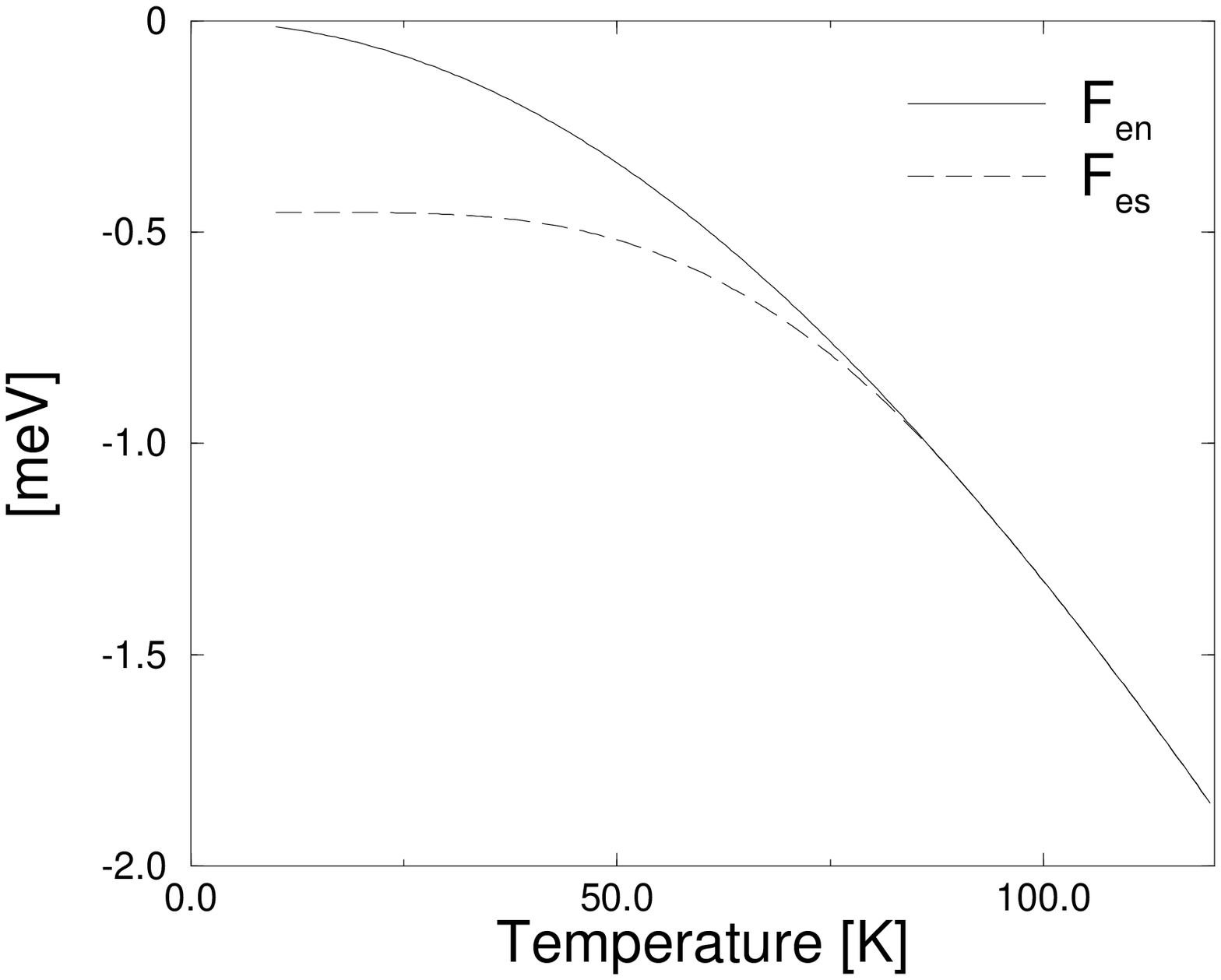,width=0.46\textwidth}}\\
\subfigure[]{\epsfig{figure=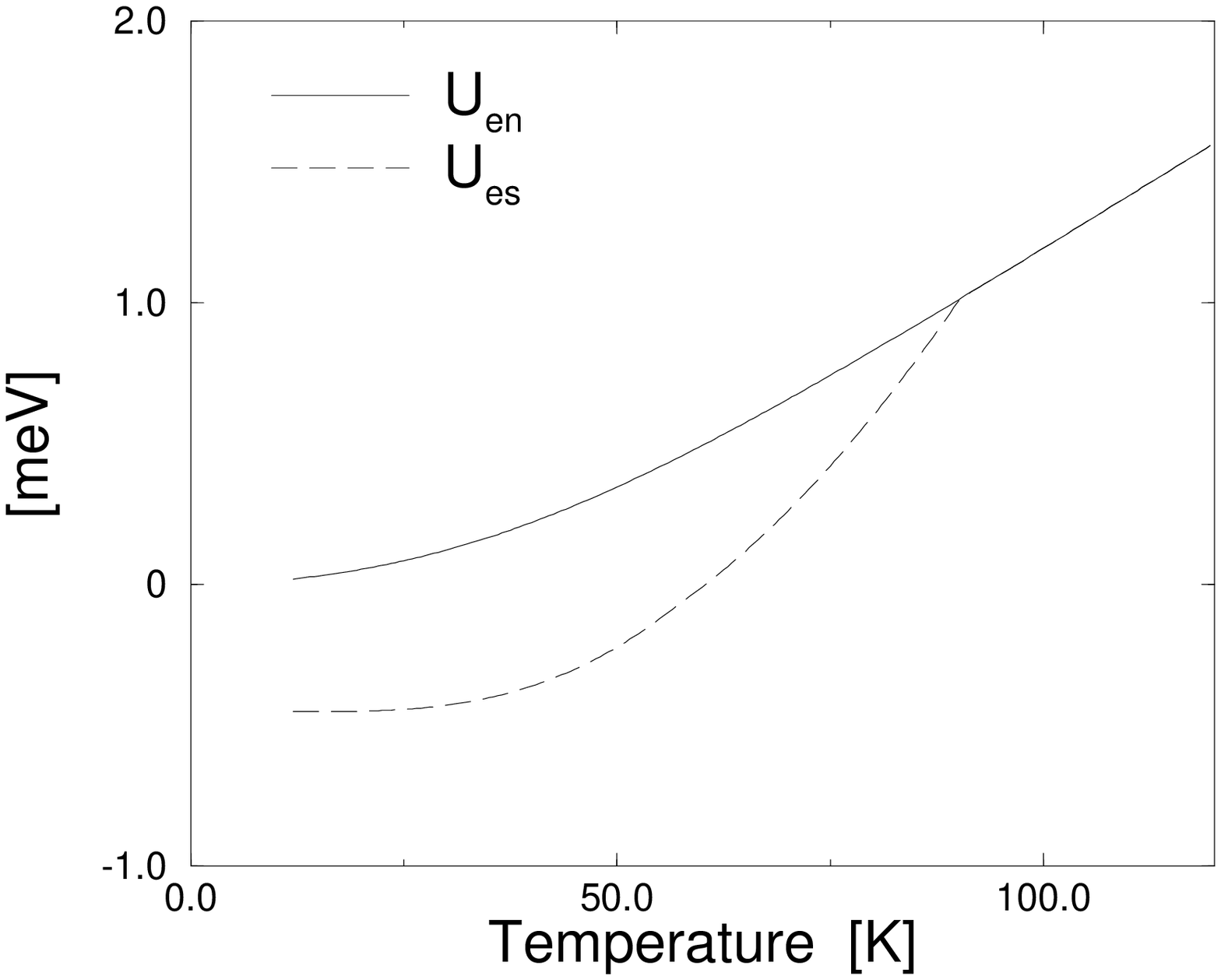,width=0.46\textwidth}}\quad
\subfigure[]{\epsfig{figure=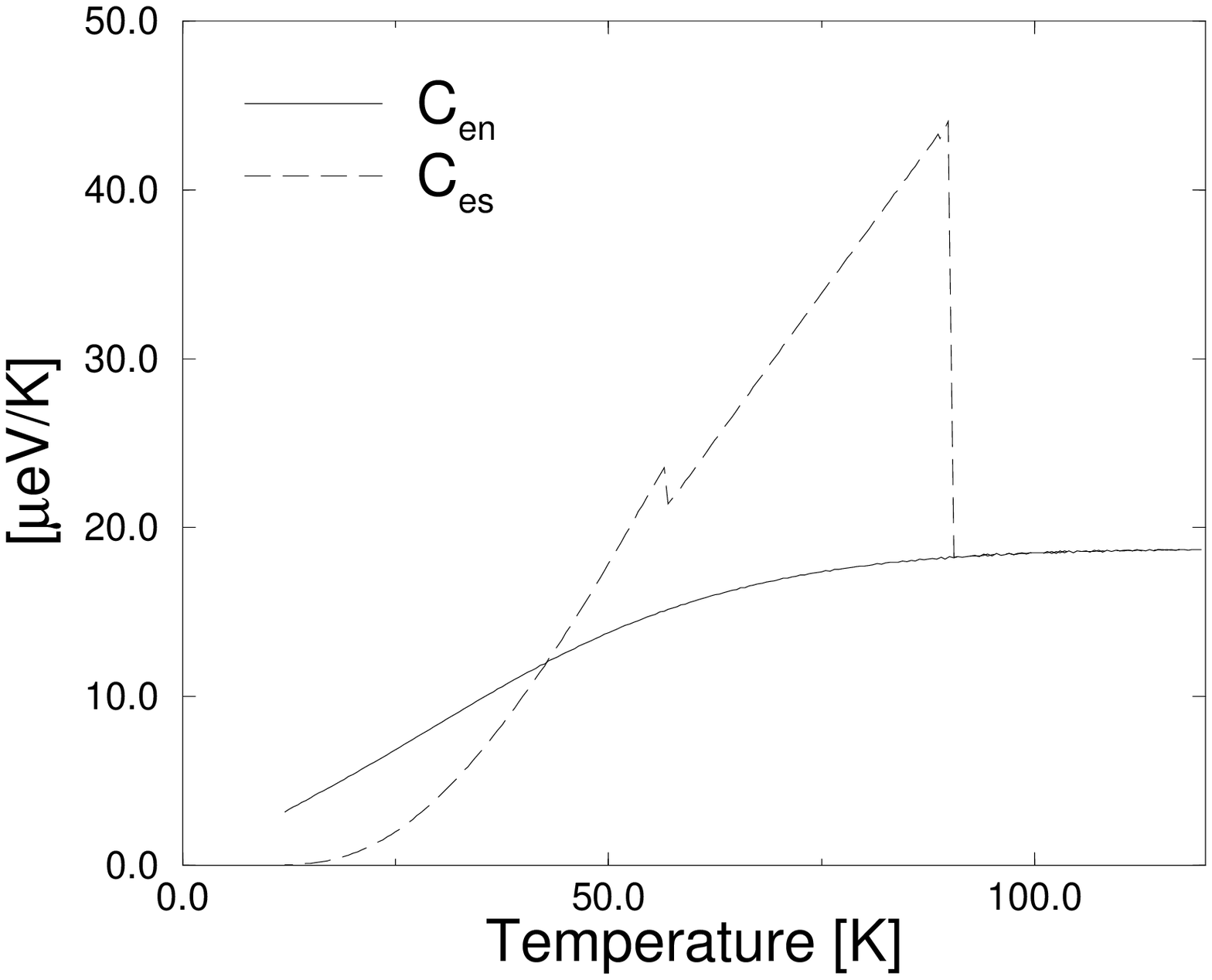,width=0.46\textwidth}}\\
\subfigure[]{\epsfig{figure=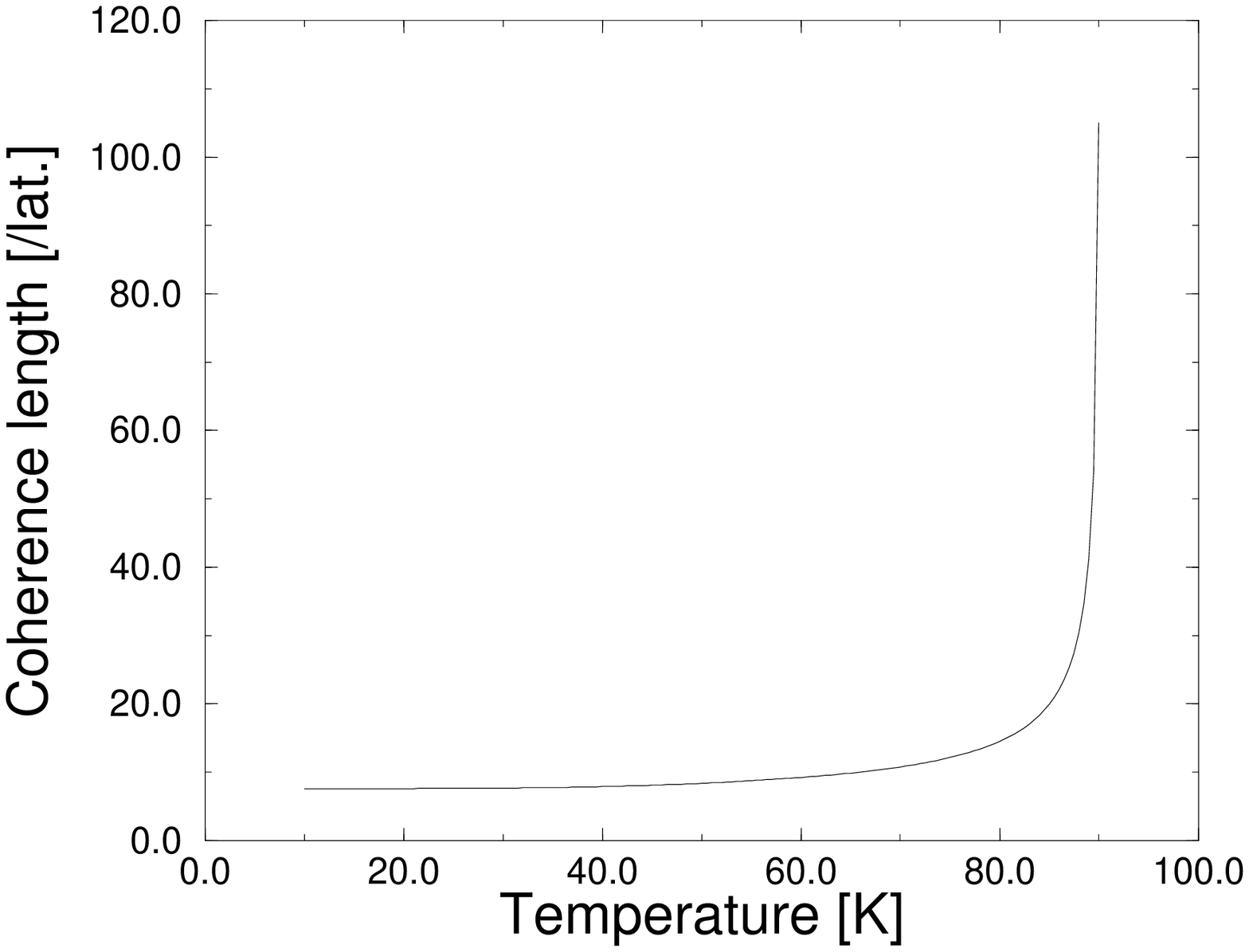,width=0.46\textwidth}}\quad
\subfigure[]{\epsfig{figure=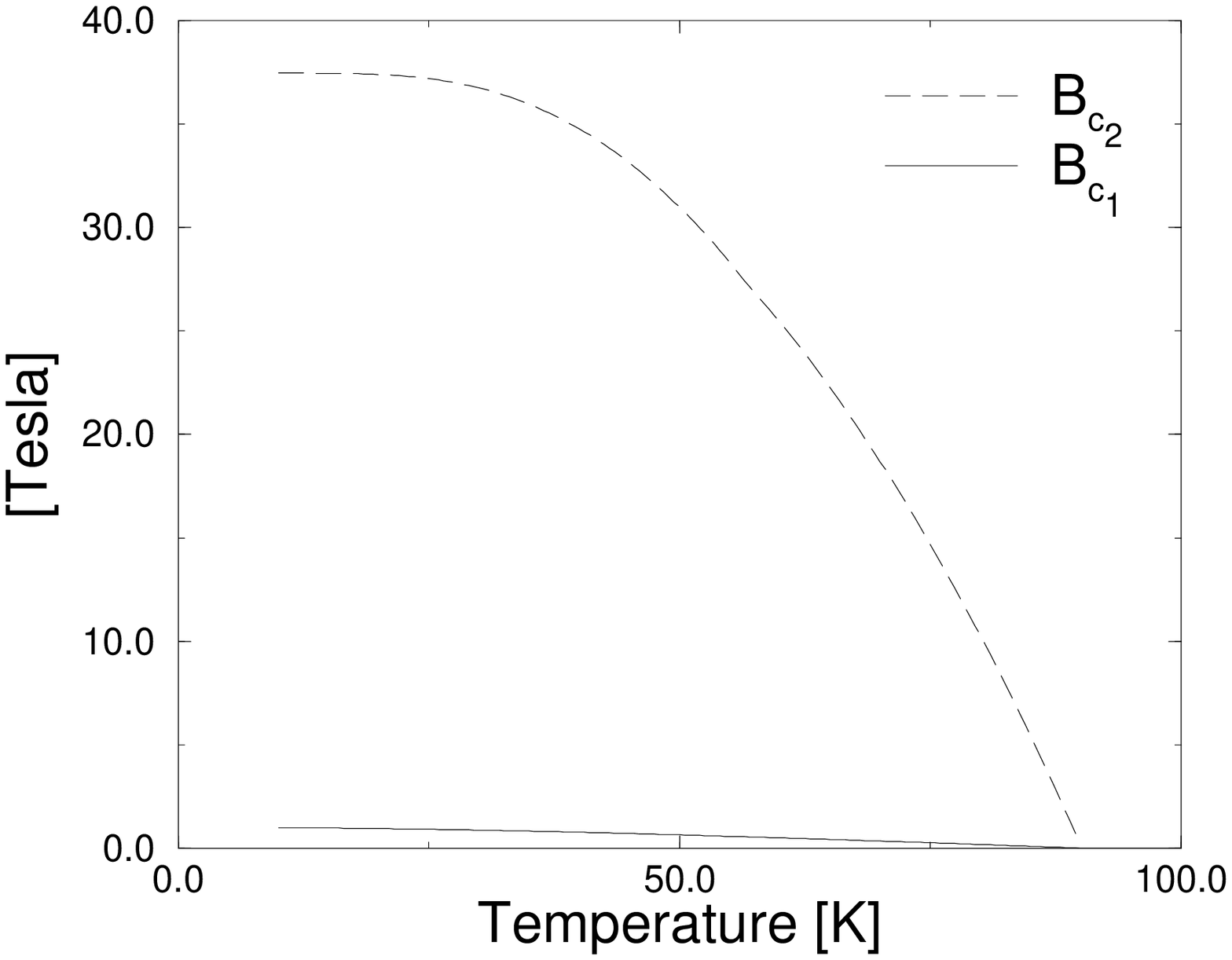,width=0.46\textwidth}}\\
\caption{}
\end{figure}
\clearpage

\enlargethispage{3cm}
\begin{figure}
\centering
\subfigure[]{\epsfig{figure=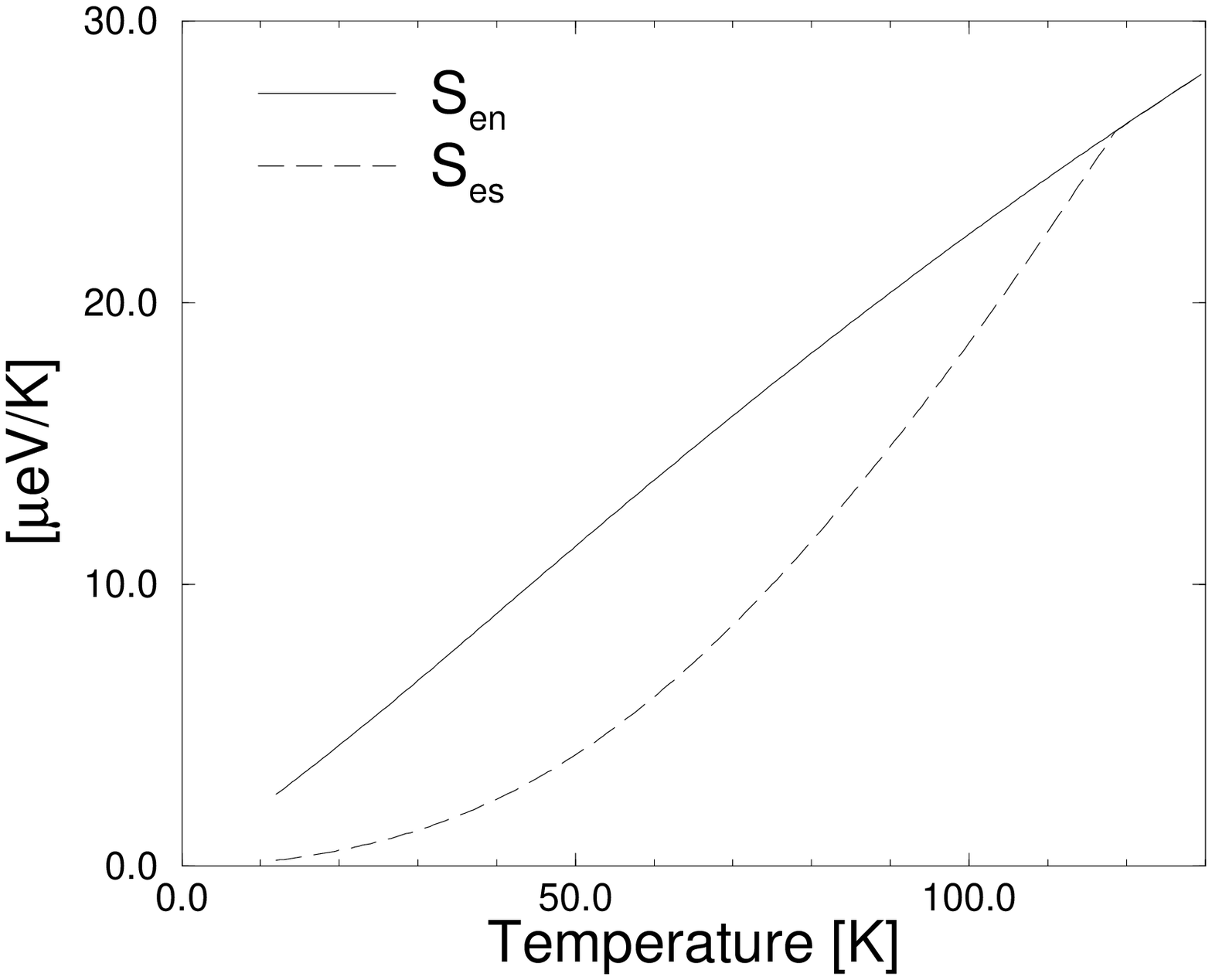,width=0.46\textwidth}}\quad
\subfigure[]{\epsfig{figure=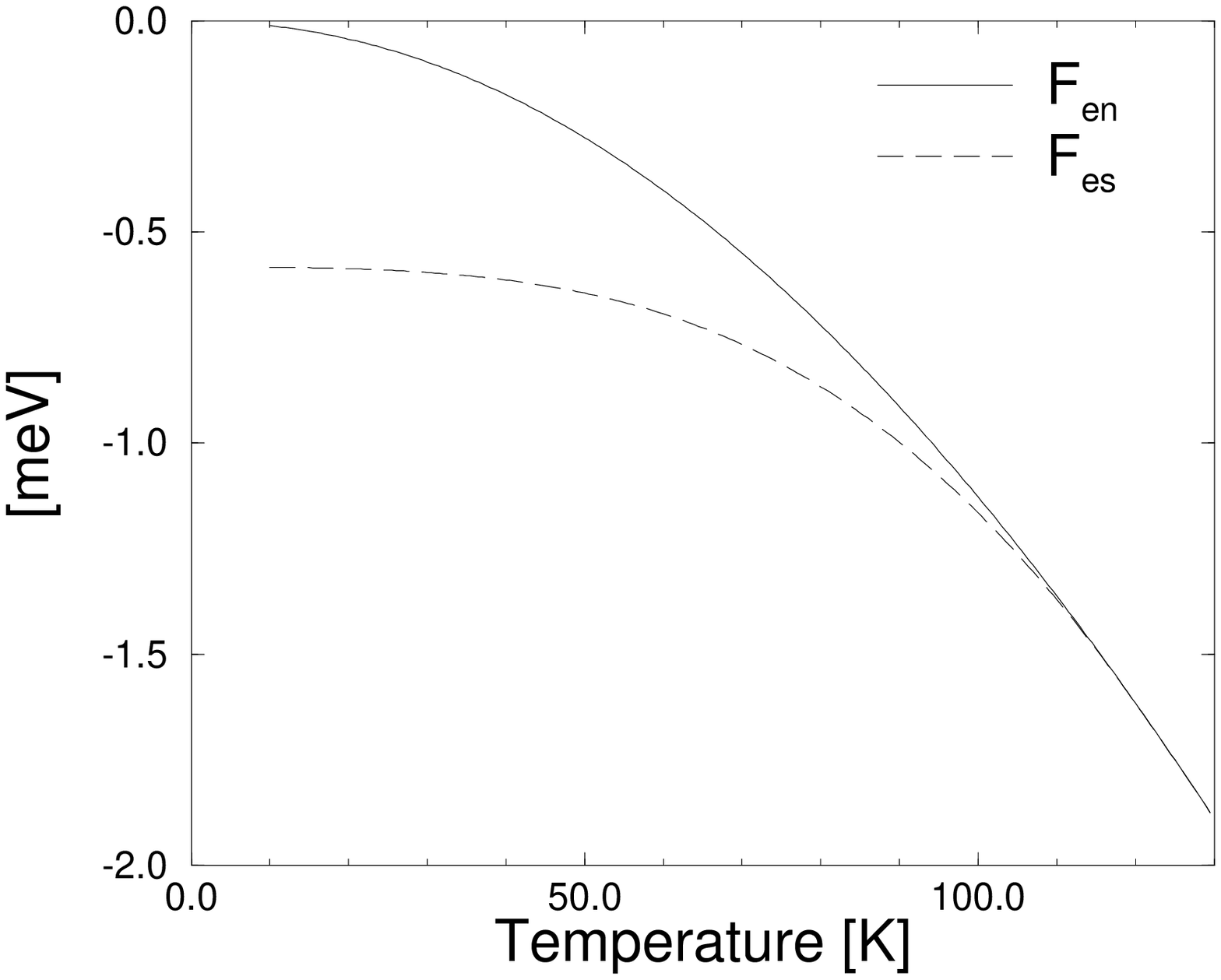,width=0.46\textwidth}}\\
\subfigure[]{\epsfig{figure=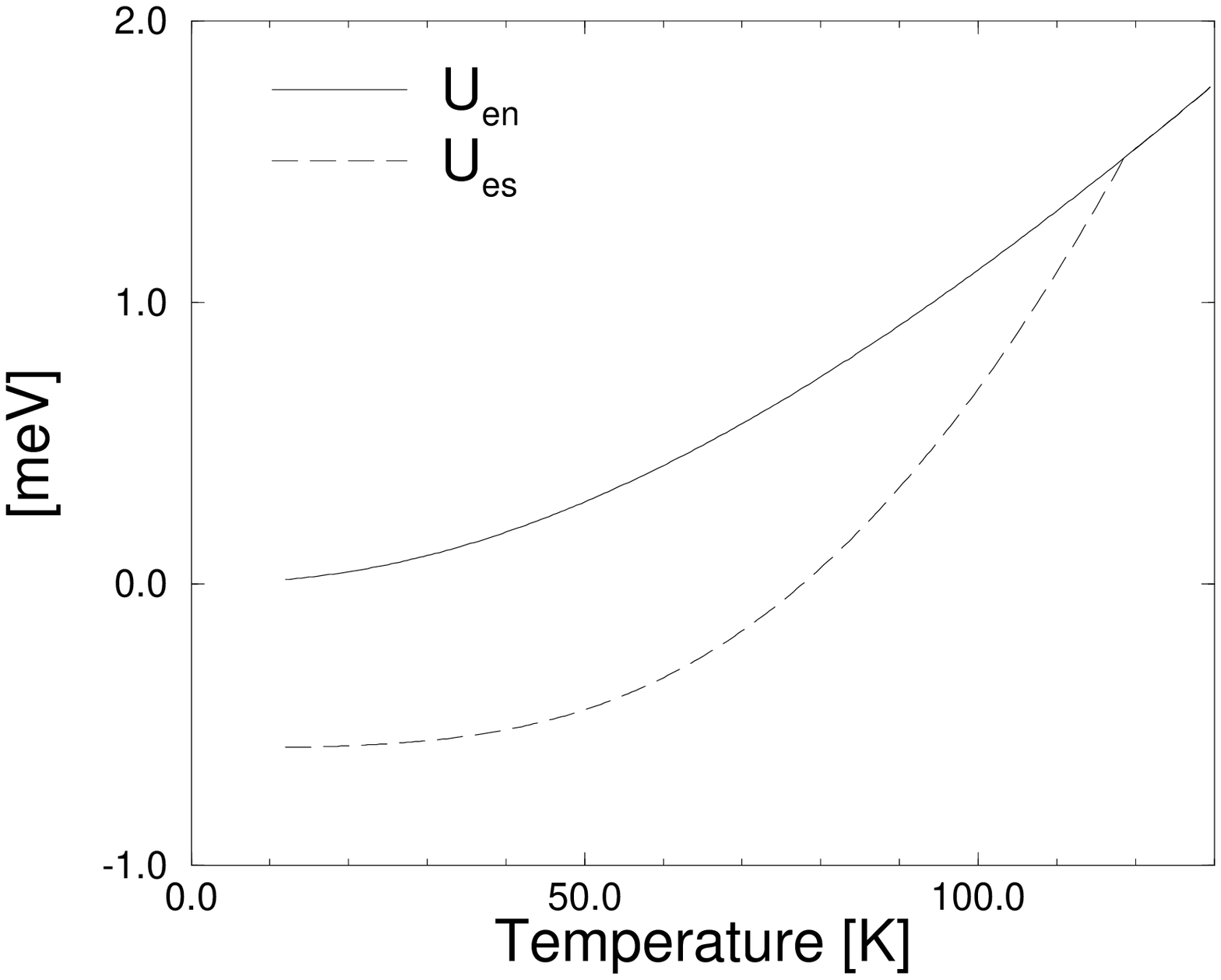,width=0.46\textwidth}}\quad
\subfigure[]{\epsfig{figure=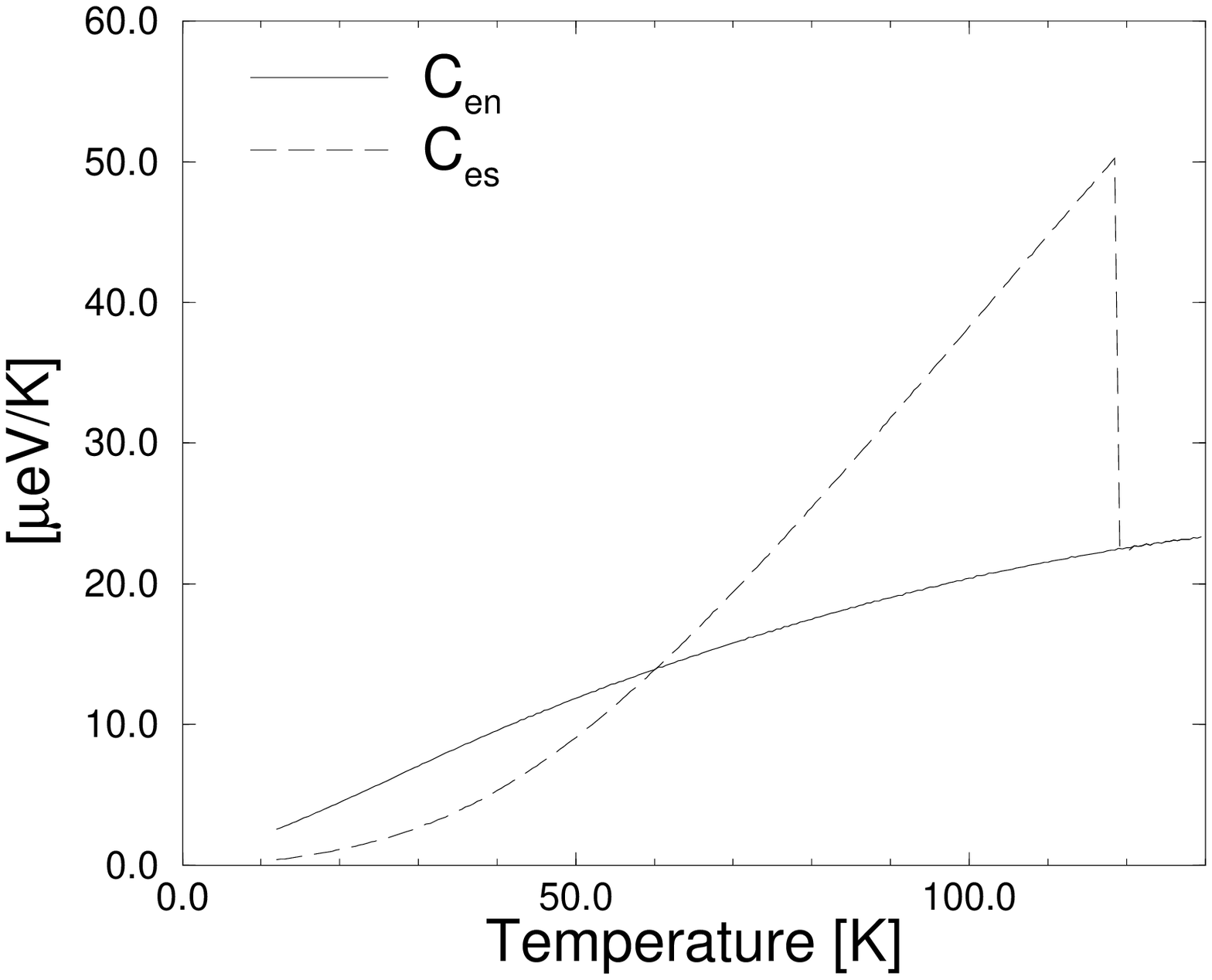,width=0.46\textwidth}}\\
\subfigure[]{\epsfig{figure=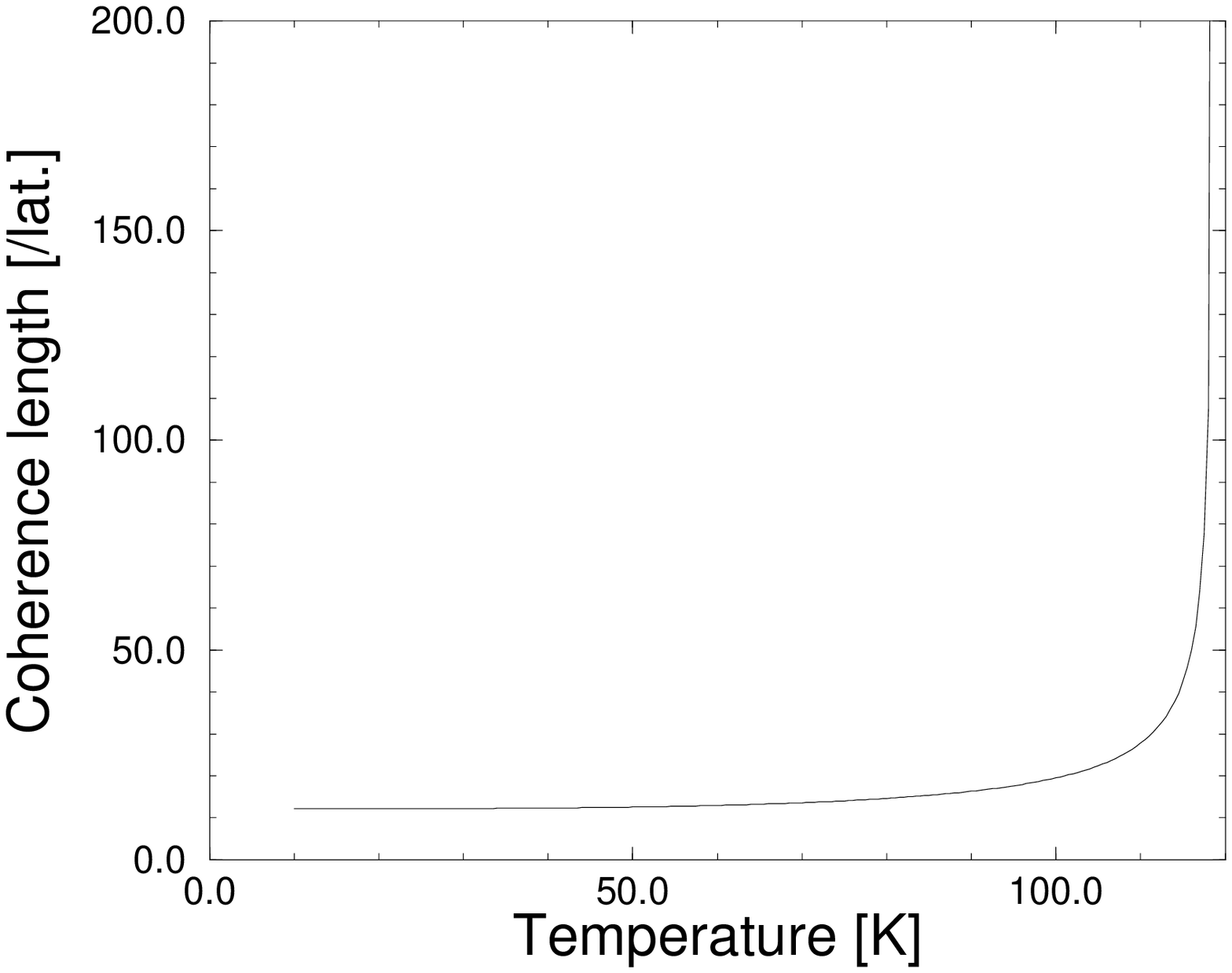,width=0.46\textwidth}}\quad
\subfigure[]{\epsfig{figure=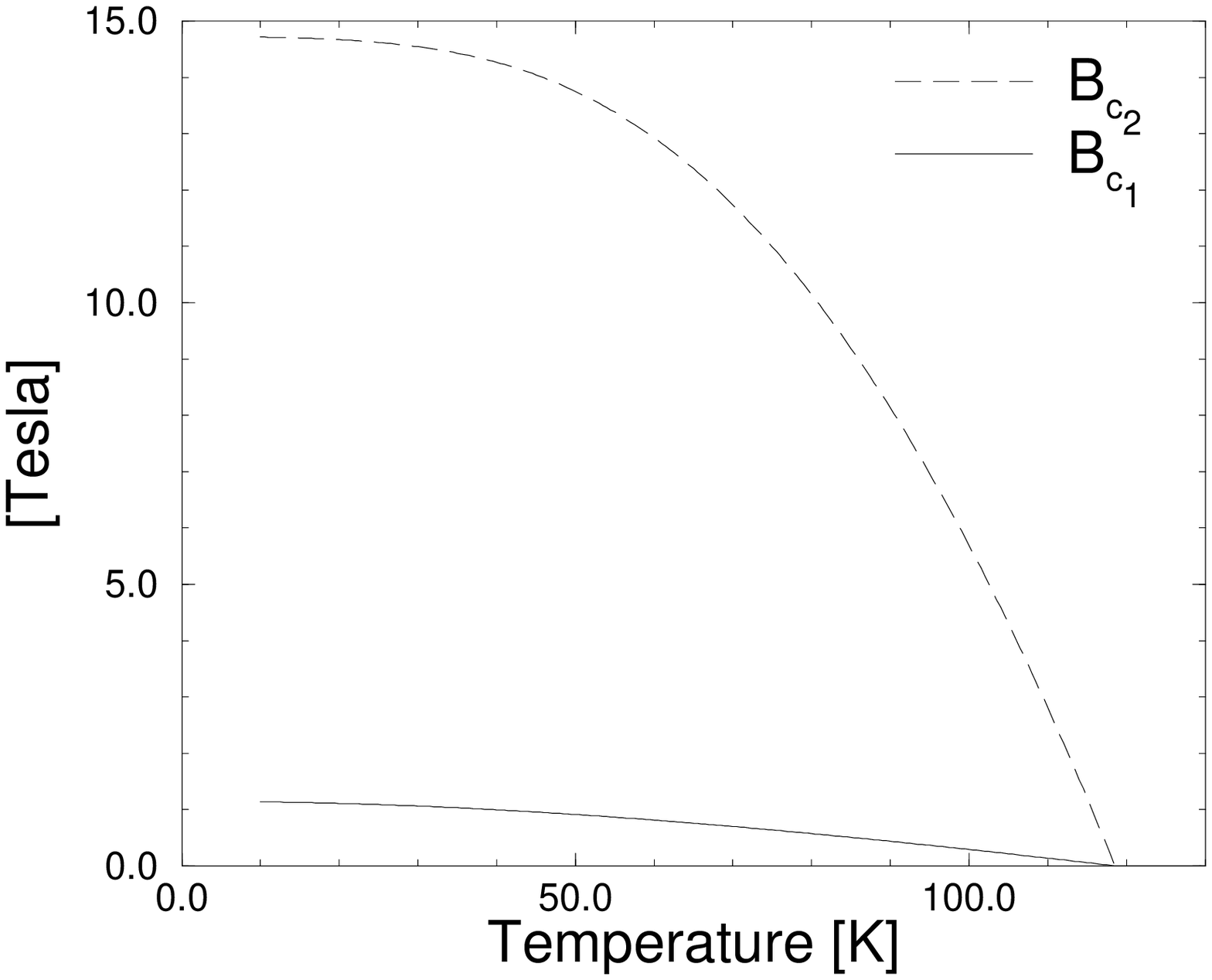,width=0.46\textwidth}}\\
\caption{}
\end{figure}
\clearpage

\begin{figure}
\enlargethispage{3cm}
\centering
\subfigure[]{\epsfig{figure=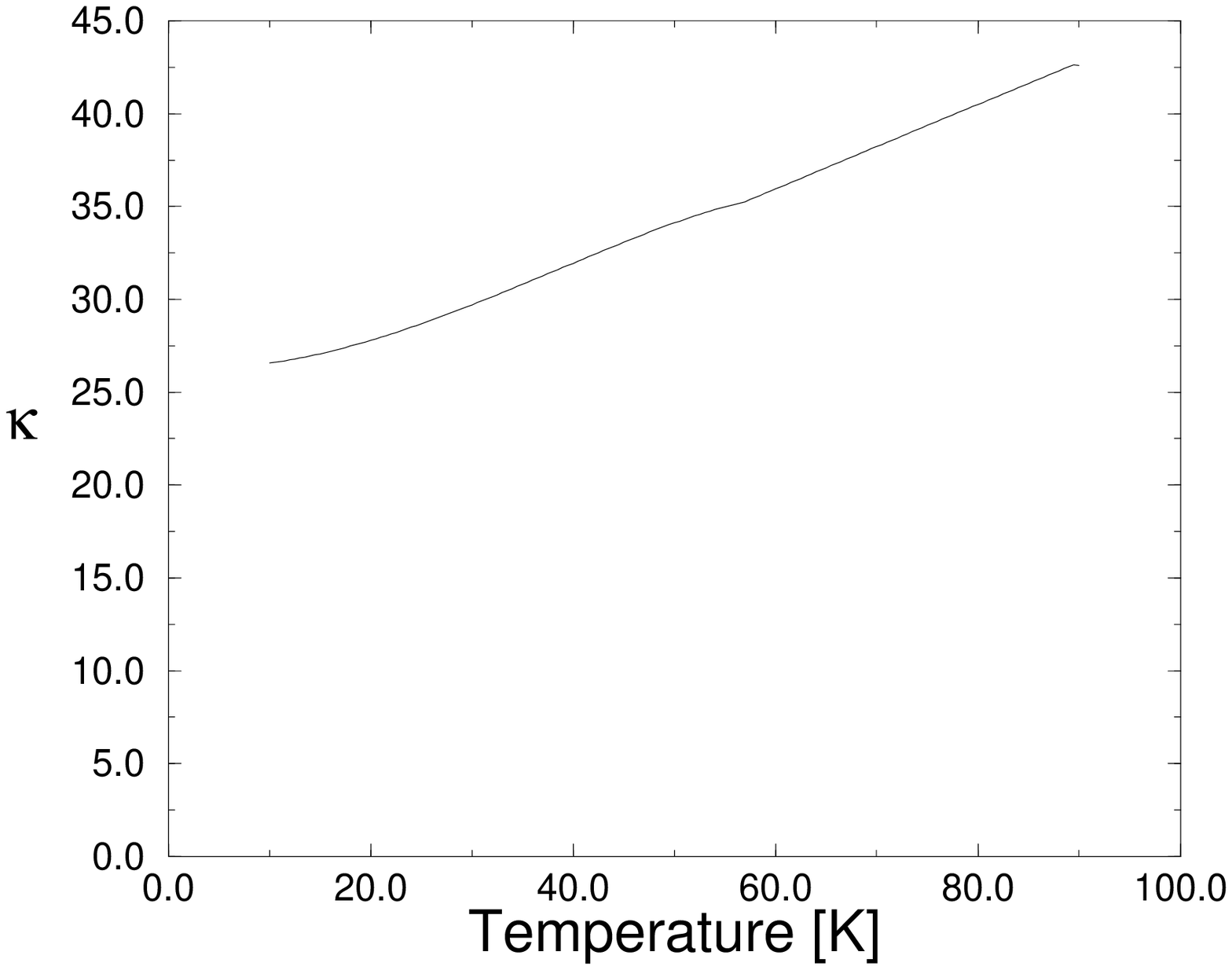,width=0.7\textwidth}}\\
\subfigure[]{\epsfig{figure=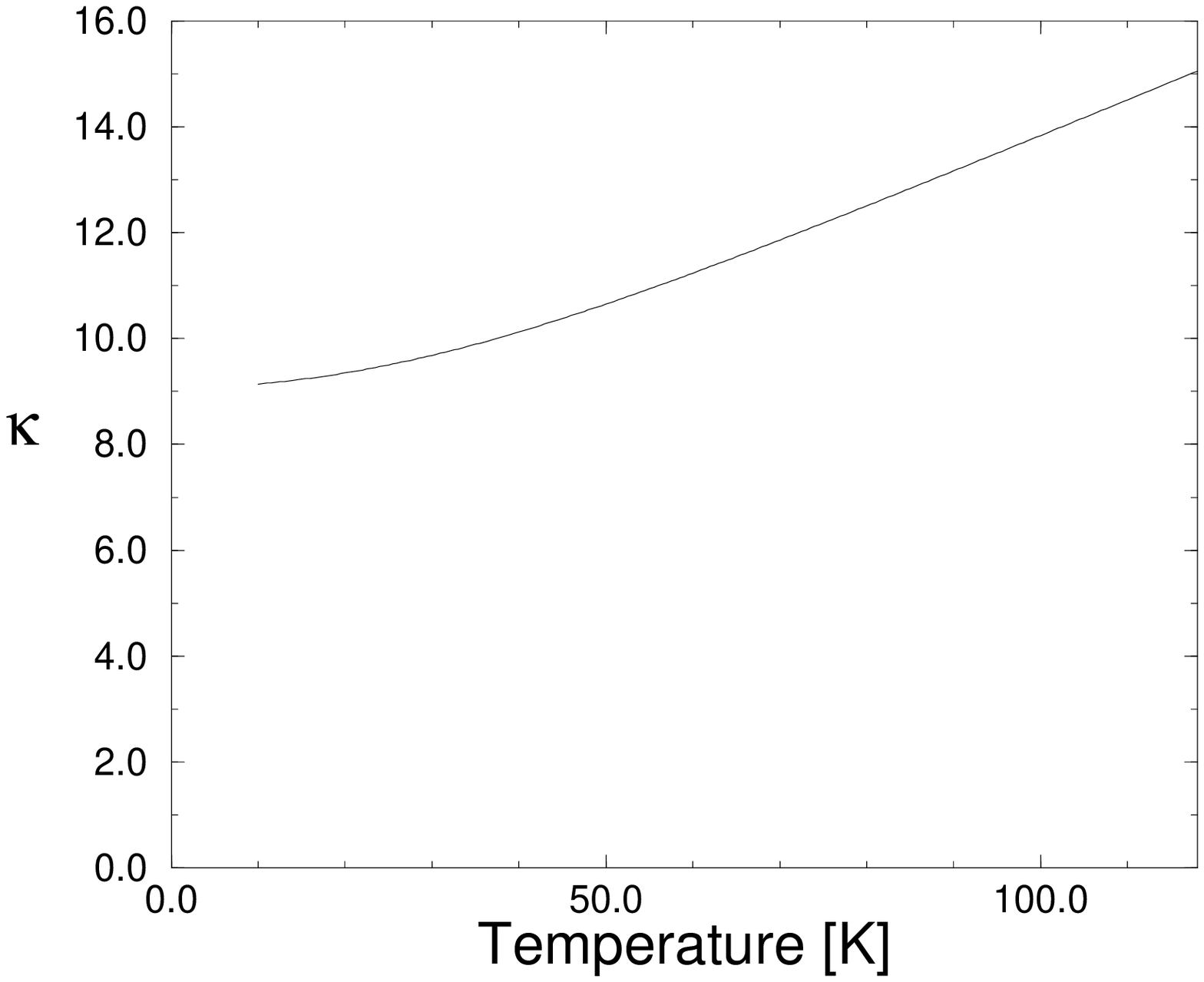,width=0.7\textwidth}}
\caption{}
\end{figure}